\documentclass[koma,10pt]{scrartcl}

\usepackage[numbers]{natbib}
\usepackage[english]{babel}
\usepackage{amsthm,amsmath,amsfonts,amssymb}
\usepackage{tikz}
\usetikzlibrary{arrows, matrix, decorations.pathmorphing, bending, calligraphy}
\usepackage{nicematrix}
\usepackage{subcaption}
\usepackage{hyperref}
\usepackage[dvipsnames]{xcolor}
\usepackage{adjustbox}

\newtheorem{theorem}{Theorem}
\newtheorem{lemma}[theorem]{Lemma}
\newtheorem{remark}{Remark}
 
\graphicspath{{../figures}}

\DeclareMathOperator*{\argmax}{arg\,max}

\author{P. Charitat$^1$ \and S. Geffray$^2$ \and C. Pouzat$^3$}
\hypersetup{
  colorlinks=true,
  linkcolor=brown,
  citecolor=blue, 
  urlcolor=orange,
  pdfauthor={Pierre Charitat, Ségolen Geffray and Christophe Pouzat},
  pdftitle={Simulation Based Inference of a Simple Neural Network Structure},
  pdfkeywords={Galves-Löcherbach Model; Monte Carlo Method; Random Graph; Point Process.},
  pdfsubject={},
  pdfcreator={}, 
  pdflang={English}
}

\title{Simulation Based Inference of a Simple Neural Network Structure}
  
\begin{document}

\maketitle

\section*{Affiliations}
IRMA\\
Université de Strasbourg and CNRS UMR 7501\\
7 rue René-Descartes\\
67084 Strasbourg Cedex\\
France\\

\noindent $^{1}$ pierre.charitat@unistra.fr \\
$^{2}$ geffray@math.unistra.fr\\
$^{3}$ Tragically, Christophe Pouzat passed away after the writing of this article. Pierre Charitat and Ségoleen Geffray would like to dedicate this work to their much esteemed friend and colleague.\\

\section*{Abstract}
Neurophysiologists are nowadays able to record from a large number of extracellular electrodes and to extract, from the raw data, the sequences of action potentials or spikes generated by many neurons. Unfortunately these ``many neurons'' still represent only a tiny fraction of the neuronal population that constitutes the network. Using association statistics such as the estimation of the cross-correlation functions, they are trying to infer the structure of the network formed by the recorded neurons. But this inference is compromised by the tremendous under-sampling of the neuronal population. We propose to focus instead on simple spike train statistics, like the empirical spikes frequency, or the interspike interval distribution. Their sampling distributions can be estimated by simulations, and, given a few observed spike train statistics, they provide enough information to infer the structure of the underlying network. We show that, on a ``toy model'', our method gives significantly better results than the sub-network reconstruction method with regards to the inference of the connection probability of the original network.

\noindent \textbf{Keywords} Galves-Löcherbach Model; Monte Carlo Method; Random Graph; Point Process.

\section{Introduction}
\label{intro}
There is an elephant in the room when the analysis of neuronal spike trains is considered. If neurophysiologists are nowadays able to record from a large number of extracellular electrodes and to extract, from the raw data, the sequences of action potentials or spikes generated by many neurons \citep{Pouzat_2016,Leblois_Pouzat_2017}; \emph{these ``many neurons'' represent only a (very) tiny fraction of the neuronal population that constitutes the network under study}. Using association statistics such as the estimation of the cross-correlation functions \citep{Perkel_Gerstein_Moore_1967, Brillinger:1976} or  stochastic intensity models \citep{Brillinger_1988,Chornoboy_Schramm_Karr_1988}, neurophysiologists try and infer the structure of the network formed by the recorded neurons. But this inference is compromised by the tremendous under-sampling of the neuronal population \citep{Moore_Et_Al_1970}. This yields a ``network picture'' usually called a \emph{functional network} \cite{sporns:2010} whose features depend strongly on the recording conditions (such as the presence/absence of a stimulation) and whose relation to the actual network structure is far from obvious.

Given what is known on the development of the brain wiring \citep[Chap. 7]{luo:2021} and \cite[Chap. 4]{sporns:2010}, we think that it is reasonable to postulate, that when a neurophysiologist \emph{repeats an experiment} on several individuals of the same species, focusing on a \emph{given network}---think of the first olfactory relay of a locust, the hippocampus of a rat, the striate cortex of a monkey---, \emph{the networks of the different individuals can be viewed as realizations of the same underlying random graph}. This is our central working hypothesis. It then makes sense---rather than reconstructing the network formed by the recorded neurons (a definitely ill-posed problem)---to focus on the \emph{generative probability distribution of the graph}, since \emph{this is what should be reproducible across experiments}. In this article, we are aiming at a proof of concept and we are going to consider a toy model for actual brain networks; a directed Erdős-Rényi model involving two parameters: the number of neurons (number of vertices) and the connection probability between any pair of neurons/vertices. Clearly, further work will have to consider more elaborate models like \emph{stochastic block models} \cite{lee.wilkinson:2019}, as well as models with motifs (that is, non independent connections) \cite{sporns:2010}.   

Now, changing the question addressed when analyzing neuronal spike trains does not lead, obviously at least, to a solution adressing the massive under-sampling problem we started with. To deal with the later, we propose to use a numerical model of \emph{the whole network}, from which we will sample a few spike trains, mimicking what is done in the real experiments. The point is that we know a lot about the morphology of the different neurons in many actual networks \cite{sporns:2010,braitenberg.shuz:1998}---this can be used to constrain the random graph model---and we also know a lot about the individual neurons physiology as well as about the physiology of their synaptic connections; that's why neurobiology textbooks \cite{luo:2021} are so thick! Assuming such a whole network simulation can be carried out satisfyingly, the problem becomes the selection of the right statistics to compute from the sampled spike trains so that the inference of the parameters of postulated random graph model is satisfying, in a sense to be precised later. In that way, we address the under-sampling problem by reproducing it numerically as faithfully as possible. This allows us to estimate the sampling distribution of our statistics, and to see if the later depend critically on the random graph parameters. This is the essence of our proposed \emph{simulation based inference} approach, and, in our case, it is directly inspired by the last example of \cite[Sec. 6, pp; 208-210]{digglegratton} and \cite{Wood_2010}. Notice also that the ``usual'' spike train analysis methods totally ignore the anatomical and physiological knowledge accumulated by decades of experimental research---a terrible waste of resources in our view, as well as a major barrier to a fruitful dialogue between experimentalists and statisticians---while our approach is capable of including such knowledge by design. 

Our numerical model follows the ``realist'' precepts of Antonio Galves, as opposed to the ``naturalist'' approach involved in detailed biophysically based models with their myriad of parameters \cite{gouwens.ea:2018}; an approach corresponding to the aphorism attributed to Einstein: ``Make a model as simple as possible, but not simpler''. Therefore, we do not model the 3D extension of actual neurons and use ``point'' neurons instead; the edges of our graphs are considered as good (enough) representations of the actual synaptic network intricacies. We consider that neurons are ``stochastic units'' and we adopt, for their dynamics, a specific Galves-Löcherbach model \cite{galves.locherbach:2013}. We invite our readers to read \cite[Chap. 1 and Appendix A]{galves.locherbach.pouzat:2024} for a critical discussion and empirical justification of these choices.

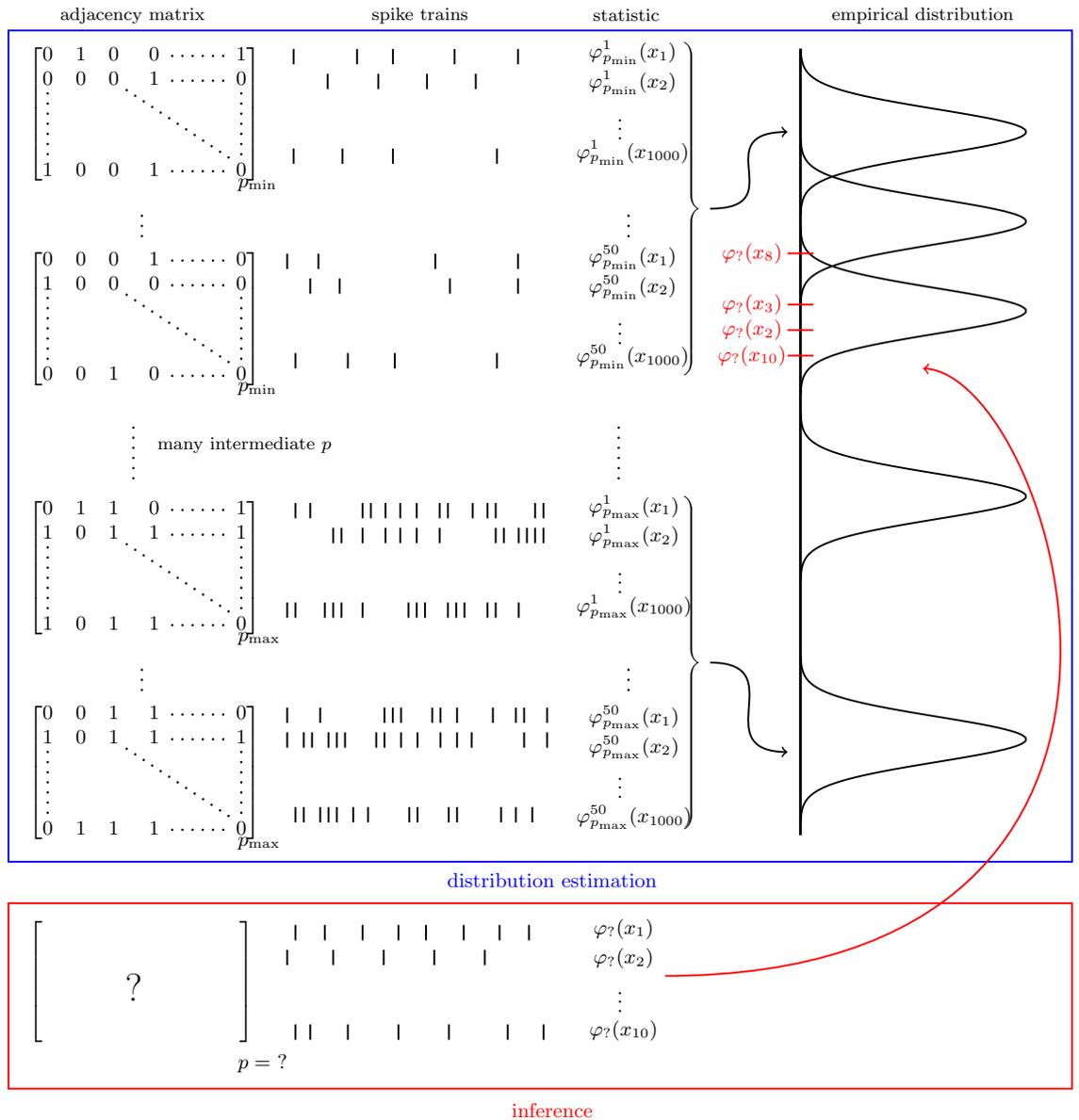
\begin{figure}[]
    \begin{adjustbox}{center}
        \centering
        \begin{tikzpicture}[
  every node/.style={font=\small}
]
\def\vertbar{\draw[thick] (0,-0.12)--(0,0.12)}
\useasboundingbox (0,0) rectangle (17, 15.35);

\scope[shift={(1.25, 16.7)},
    transform canvas={scale=0.9}]
% - TOP TEXT ---------------------------------
\begin{scope}[yshift=15]
\node at (1.74, 0) {\footnotesize adjacency matrix};
\node at (6.2, 0) {\footnotesize spike trains};
\node at (9.4, 0) {\footnotesize statistic};
\node at (14, 0) {\footnotesize empirical distribution};
\end{scope}
% - TOP TEXT ---------------------------------

\begin{scope}[blue]
\node[rectangle, draw, thick, 
        anchor=north west,
        minimum width=16.5cm,
        minimum height=13.4cm] at (-0.2, 0.3) {};
\node at (8.25, -13.4) {distribution estimation};
\end{scope}

% - PMIN -------------------------------------
\begin{scope}
% ----- ROW 1 --------------------------------
\begin{scope}[shift={(0,0.2)}]
% ----------- ADJACENCY MATRIX ---------------
\node[anchor=north west] at (0,0) 
{
$\NiceMatrixOptions{xdots/shorten=0.5em}
\begin{bNiceMatrix}
  0 & 1 & 0 & 0  & \Cdots & 1 \\
  0 & 0 & 0 & 1 & \Cdots & 0 \\
  \Vdots & & & \Ddots & & \Vdots \\
  & & & & & \\
  1 & 0 & 0 & 1 & \Cdots & 0
\end{bNiceMatrix}$
};
\node[anchor=north west] at (3.25, -2.25) {$p_{\min}$};
% ----------- ADJACENCY MATRIX ---------------

% ----------- SPIKE TRAIN ---------------
\matrix [anchor=north west,column sep=1mm] (st1) at (4, -0.065) 
{
     & \vertbar; & & & & & & & & & \vertbar; & & & & & \vertbar; & & & & & & & & & \vertbar; & & & & & & & & & \vertbar; \\[1.2mm]
    & & & & & &  \vertbar; & & & & & & & \vertbar; & & & & & & &  \vertbar; & & & & & & &  \vertbar; & & & & & & \\
    \\[9mm]
    & \vertbar; & & & & & & & \vertbar; & & & & & & &  \vertbar; & & & & & & & & & & & & & & & \vertbar;  & &\\
};
% ----------- SPIKE TRAIN ---------------

% ----------- STATISTIC ---------------
\node[anchor=north west] at (8.5, 0.05) {%
$\begin{matrix}
    \varphi^1_{p_{\min}}(x_1) \\[0.75mm] 
    \varphi^1_{p_{\min}}(x_2) \\
     \\[3.25mm]
    \varphi^1_{p_{\min}}(x_{1000})
\end{matrix}$};
\node at (9.3, -1.35) {$\vdots$};
% ----------- STATISTIC ---------------
\end{scope}
% ----- ROW 1 --------------------------------

% ----------- VDOTS ---------------
\node at (1.75, -2.65) {$\begin{NiceMatrix} \Vdots \end{NiceMatrix}$};
\node at (9.3, -2.65) {$\begin{NiceMatrix} \Vdots \end{NiceMatrix}$};
% ----------- VDOTS ---------------

% ----- ROW 50 -------------------------------
\begin{scope}[shift={(0, -3.15)}]
% ----------- ADJACENCY MATRIX ---------------
\node[anchor=north west] at (0,0) 
{
$\NiceMatrixOptions{xdots/shorten=0.5em}
\begin{bNiceMatrix}
  0 & 0 & 0 & 1  & \Cdots & 0 \\
  1 & 0 & 0 & 0 & \Cdots & 0 \\
  \Vdots & & & \Ddots & & \Vdots \\
  & & & & & \\
  0 & 0 & 1 & 0 & \Cdots & 0
\end{bNiceMatrix}$
};
\node[anchor=north west] at (3.25, -2.25) {$p_{\min}$};
% ----------- ADJACENCY MATRIX ---------------

% ----------- SPIKE TRAIN ---------------
\matrix [anchor=north west,column sep=1mm] at (4, -0.065)
{
     \vertbar; & & & & \vertbar; & & & & & & & & & & & & & & & & & \vertbar;  & & & & & & & & & & & & \vertbar; \\[1.2mm]
    & & & \vertbar; & & & &  \vertbar; & & & & & & & & & & & & & & & & \vertbar;  & & & &  & & & & & & \vertbar; \\
    \\[9mm]
    & \vertbar; & & & & & & & \vertbar; & & & & & & &  \vertbar; & & & & & & & & & & & & & & & \vertbar;  & &\\
};
% ----------- SPIKE TRAIN ---------------

% ----------- STATISTIC ---------------
\node[anchor=north west] at (8.5, 0.05) {%
$\begin{matrix}
    \varphi^{50}_{p_{\min}}(x_1) \\[0.75mm] 
    \varphi^{50}_{p_{\min}}(x_2) \\
     \\[3.25mm]
    \varphi^{50}_{p_{\min}}(x_{1000})
\end{matrix}$};
\node at (9.3, -1.35) {$\vdots$};
% ----------- STATISTIC ----------------------
\end{scope}
% ----- ROW 50 -------------------------------

% ----- BRACE --------------------------------
\begin{scope}[shift={(10.4,0.1)} ]
\draw[decorate, decoration={calligraphic brace, amplitude=6pt}, thick] 
    (0, 0) -- (0,-5.35)
  node[midway, right=8pt] {};
\end{scope}
% ----- BRACE --------------------------------
\end{scope}
% - PMIN -------------------------------------

% - VDOTS ------------------------------------
\node at (1.67, -6.35) {$\begin{NiceMatrix}&  \Vdots \\ & \\ \end{NiceMatrix}$};
\node at (9.2, -6.35) {$\begin{NiceMatrix}& \Vdots \\ & \\ \end{NiceMatrix}$};

\node[anchor=west] at (2, -6.35) {\footnotesize many intermediate $p$};
% - VDOTS ------------------------------------

% - PMAX -------------------------------------
\begin{scope}[shift={(0,-7.25)}]
% ----- ROW 1 --------------------------------
\begin{scope}[shift={(0,0.2)}]
% ----------- ADJACENCY MATRIX ---------------
\node[anchor=north west] at (0,0) 
{
$\NiceMatrixOptions{xdots/shorten=0.5em}
\begin{bNiceMatrix}
  0 & 1 & 1 & 0  & \Cdots & 1 \\
  1 & 0 & 1 & 1 & \Cdots & 1 \\
  \Vdots & & & \Ddots & & \Vdots \\
  & & & & & \\
  1 & 0 & 1 & 1 & \Cdots & 0
\end{bNiceMatrix}$
};
\node[anchor=north west] at (3.25, -2.25) {$p_{\max}$};
% ----------- ADJACENCY MATRIX ---------------

% ----------- SPIKE TRAIN ---------------
\matrix [anchor=north west,column sep=1mm] (st1) at (4, -0.065) 
{
     & \vertbar; & & \vertbar; & & & & & & & \vertbar; & \vertbar; & & \vertbar; & & \vertbar; & & \vertbar; & & & \vertbar;& \vertbar;& & & \vertbar; & & \vertbar;& \vertbar; & & & & & \vertbar; & \vertbar; \\[1.2mm]
    & & & & & &  \vertbar; & \vertbar; & & & \vertbar; & & & \vertbar; & & \vertbar; & & \vertbar; & & &  \vertbar; & & & & & & &  \vertbar; & \vertbar;& & \vertbar;& \vertbar;& \vertbar;&\vertbar; \\
    \\[9mm]
    \vertbar; & \vertbar; & & & & \vertbar;& \vertbar;& \vertbar;& & & \vertbar;& & & & & & \vertbar; & \vertbar; & \vertbar; & & & \vertbar;&\vertbar; &\vertbar; & & &\vertbar; & \vertbar; & & & \vertbar;  & &\\
};
% ----------- SPIKE TRAIN ---------------

% ----------- STATISTIC ---------------
\node[anchor=north west] at (8.5, 0.05) {%
$\begin{matrix}
    \varphi^1_{p_{\max}}(x_1) \\[0.75mm] 
    \varphi^1_{p_{\max}}(x_2) \\
     \\[3.25mm]
    \varphi^1_{p_{\max}}(x_{1000})
\end{matrix}$};
\node at (9.3, -1.35) {$\vdots$};
% ----------- STATISTIC ---------------
\end{scope}
% ----- ROW 1 --------------------------------

% ----------- VDOTS ---------------
\node at (1.75, -2.65) {$\begin{NiceMatrix} \Vdots \end{NiceMatrix}$};
\node at (9.3, -2.65) {$\begin{NiceMatrix} \Vdots \end{NiceMatrix}$};

% ----------- VDOTS ---------------

% ----- ROW 50 -------------------------------
\begin{scope}[shift={(0, -3.15)}]
% ----------- ADJACENCY MATRIX ---------------
\node[anchor=north west] at (0,0) 
{
$\NiceMatrixOptions{xdots/shorten=0.5em}
\begin{bNiceMatrix}
  0 & 0 & 1 & 1  & \Cdots & 0 \\
  1 & 0 & 1 & 1 & \Cdots & 1 \\
  \Vdots & & & \Ddots & & \Vdots \\
  & & & & & \\
  0 & 1 & 1 & 1 & \Cdots & 0
\end{bNiceMatrix}$
};
\node[anchor=north west] at (3.25, -2.25) {$p_{\max}$};
% ----------- ADJACENCY MATRIX ---------------

% ----------- SPIKE TRAIN ---------------
\matrix [anchor=north west,column sep=1mm] at (4, -0.065) 
{
     \vertbar; & & & & \vertbar; & & & & & & & & \vertbar;& \vertbar;& \vertbar;& & & & \vertbar;& \vertbar;& & \vertbar;  & & & & & \vertbar;& & & \vertbar;& \vertbar;& & & \vertbar; \\[1.2mm]
    \vertbar;& &\vertbar; & \vertbar; & & \vertbar;& \vertbar;&  \vertbar; & & & & \vertbar;& \vertbar;& & \vertbar;& & \vertbar;& & & \vertbar;& &\vertbar; & & \vertbar;  & & & &  & & & \vertbar;& & & \vertbar; \\
    \\[9mm]
    & \vertbar; & \vertbar;& &\vertbar; &\vertbar; & \vertbar;& & \vertbar; & & \vertbar;& & & & &  \vertbar; & \vertbar;& & & &\vertbar; &\vertbar; & & & & & & \vertbar;& &\vertbar; &  & \vertbar;&\\
};
% ----------- SPIKE TRAIN ---------------

% ----------- STATISTIC ---------------
\node[anchor=north west] at (8.5, -0.05) {%
$\begin{matrix}
    \varphi^{50}_{p_{\max}}(x_1) \\[0.75mm] 
    \varphi^{50}_{p_{\max}}(x_2) \\
     \\[3.25mm]
    \varphi^{50}_{p_{\max}}(x_{1000})
\end{matrix}$};
\node at (9.3, -1.35) {$\vdots$};
% ----------- STATISTIC ----------------------
\end{scope}
% ----- ROW 50 -------------------------------

% ----- BRACE --------------------------------
\begin{scope}[shift={(10.4,0.2)} ]
\draw[decorate, decoration={calligraphic brace, amplitude=6pt}, thick] 
    (0, 0) -- (0,-5.5)
  node[midway, right=8pt] {};
\end{scope}
% ----- BRACE --------------------------------
\end{scope}
% - PMAX -------------------------------------

% ----- ROW ? --------------------------------
\begin{scope}[shift={(0,-13.9)}]
% ----------- ADJACENCY MATRIX ---------------
\node[anchor=north west] at (0,0) 
{
$\NiceMatrixOptions{xdots/shorten=0.5em}
\begin{bNiceMatrix}
  \phantom{0} & \phantom{0} & \phantom{0} & \phantom{0} & \phantom{000} & \phantom{0} \\
  & & & & & \\
  & & & & & \\
  & & & & & \\
  & & & & & \\
\end{bNiceMatrix}$
};
\node[anchor=north west] at (1.5, -0.85) {\LARGE?};
\node[anchor=north west] at (3.25, -2.1) {$p =\ ?$};
% ----------- ADJACENCY MATRIX ---------------

% ----------- SPIKE TRAIN --------------------
\matrix [anchor=north west,column sep=1mm] (st1) at (4, -0.065) 
{
     & \vertbar; & & & & \vertbar;& & & & & \vertbar; & & & & & \vertbar; & & & & \vertbar;& & & & & \vertbar; & & & & & \vertbar;& & & & \vertbar; \\[1.2mm]
    \vertbar; & & & & & &  \vertbar; & & & & & & & \vertbar; & & & & & & &  \vertbar; & & & & & & &  \vertbar; & & & & & & \\
    \\[9mm]
    & \vertbar; & &\vertbar; & & & & & \vertbar; & & & & & & &  \vertbar; & & & & & & & \vertbar;& & & & & & & & \vertbar;  & & & & &\vertbar;\\
};
% ----------- SPIKE TRAIN --------------------

% ----------- STATISTIC ----------------------
\node[anchor=north west] at (8.7, 0.05) {%
$\begin{matrix}
    \varphi_{\text{?}}(x_1) \\[0.75mm] 
    \varphi_{\text{?}}(x_2) \\
     \\[3.5mm]
    \varphi_{\text{?}}(x_{10})
\end{matrix}$};
\node at (9.3, -1.3) {$\vdots$};
% ----------- STATISTIC ----------------------

% ----------- RECTANGLE ----------------------
\begin{scope}[red]
\node[rectangle, draw, thick, 
        anchor=north west,
        minimum width=16.5cm,
        minimum height=2.9cm] at (-0.2, 0.15) {};
\node at (8.25, -3.1) {inference};
\end{scope}
% ----------- RECTANGLE ----------------------

\end{scope}
% - ROW ? ------------------------------------

% - GRAPH ------------------------------------
\begin{scope}[shift={(12.1,0)}]
    \draw[very thick] (0, 0)--(0,-12.3);
    \draw[shift={(0,-1.3)}, rotate=-90, thick] plot[domain=-1.3:1.3,samples=50, smooth] (\x, {3.5*exp(-4*\x*\x)}) node[pos=] (gaussian1) {};
    \draw[shift={(0,-2.7)}, rotate=-90, thick] plot[domain=-1.3:1.3,samples=50, smooth] (\x, {3.5*exp(-4*\x*\x)}) node[pos=] (gaussian2) {};
    \draw[shift={(0,-4.1)}, rotate=-90, thick] plot[domain=-1.3:1.3,samples=50, smooth] (\x, {3.5*exp(-4*\x*\x)}) node[pos=] (gaussian3) {};
    \draw[shift={(0,-7)}, rotate=-90, thick] plot[domain=-1.3:1.3,samples=50, smooth] (\x, {3.5*exp(-4*\x*\x)}) node[pos=] (gaussian4) {};
    \draw[shift={(0,-10.8)}, rotate=-90, thick] plot[domain=-1.3:1.3,samples=50, smooth] (\x, {3.5*exp(-4*\x*\x)}) node[pos=] (gaussian5) {};

    % - SAMPLES ------------------------------
    \begin{scope}[shift={(-0.2,-3.2)}]
        \draw[thick,color=red] (0, 0) -- (0.4, 0);
        \node[color=red,scale=1] at (-0.55, 0) {$\varphi_{\text{?}}(x_8)$};
    \end{scope}

    \begin{scope}[shift={(-0.2,-4)}]
        \draw[thick,color=red] (0, 0) -- (0.4, 0);
        \node[color=red,scale=1] at (-0.55, 0) {$\varphi_{\text{?}}(x_3)$};
    \end{scope}

    \begin{scope}[shift={(-0.2,-4.4)}]
        \draw[thick,color=red] (0, 0) -- (0.4, 0);
        \node[color=red,scale=1] at (-0.55, 0) {$\varphi_{\text{?}}(x_2)$};
    \end{scope}

    \begin{scope}[shift={(-0.2,-4.8)}]
        \draw[thick,color=red] (0, 0) -- (0.4, 0);
        \node[color=red,scale=1] at (-0.55, 0) {$\varphi_{\text{?}}(x_{10})$};
    \end{scope}
    % - SAMPLES ------------------------------

\end{scope}

% - GRAPH ------------------------------------

% - ARROWS -----------------------------------
\draw[->, thick] (10.7, -2.6) .. controls (12, -2.5) and (10.6,-1.3) .. (11.9, -1.3);
\draw[->, thick] (10.7, -9.75) .. controls (12, -9.6) and (10.6,-11) .. (11.9, -11);
\draw[->,thick, color=red] (10, -15.1) .. controls (19, -14.5) and (16, -5) .. (14, -5);
% - ARROWS -----------------------------------
\endscope
\end{tikzpicture}
\vspace{1cm}
\end{adjustbox}
    \caption{\label{fig:summary}First column: Graphs are represented by their \emph{non-symmetric} adjacency matrices. They are realizations of a \emph{directed} Erdős-Renyi model with 1000 vertices and a connection probability $p$. All rows within the blue box: for each $p$ value, 50 realizations are considered. Row within the red box: actual data are assumed to originate from a graph realization with unknown $p$. Our problem is to make inference on $p$.
    Second column: Rows within the blue box: the neurons making the vertices of the graphs are all described by a Galves-Löcherbach model and the spike trains---realizations of a multivariate point process---are simulated; when $p$ is small ($p_{min}$), neurons generate few spikes; when $p$ is large ($p_{max}$), they generate many spikes. Row within the red box: the spike trains of 10 randomly selected neurons are observed in actual experiment and that is all that is seen from the network.
    Third column: a simple statistic---\emph{e.g.}, the spike frequency---is computed from the simulated (blue box) and observed (red box) spike trains.
    Fourth column: Blue box: for each $p$ value, the sampling distribution of the statistic is estimated. Red box and red arrow, the estimated sampling distributions that best matches the observed statistics is found and provide our estimate $\hat{p}$ of $p$. }
\end{figure}

The article is organized as follows. Sec.~\ref{sec:data-generation-model} describes the random graph model (Sec.~\ref{sec:directed-ER-model}), then the neural dynamics (Sec.~\ref{sec:basic-discrete-time-model}-\ref{sec:dynamics}), before specifying what are known parameters with what values and what is unknown with values to be estimated from the data (Sec.~\ref{sec:simple-setting-considered-in-this-article}). Sec.~\ref{sec:general_method} describes the general simulation-based inference approach used in this article. The first part (Sec.~\ref{sec:an-intractable-likelihood}-\ref{general_method:particular_case}) explain why a classical maximum likelihood inference cannot be implemented, leading to the use of the sampling distribution of a ``well chosen'' summary statistic. Sec.~\ref{model:sbi} makes explicit the inference part of the approach. Sec.~\ref{sec:estimating_likelihood} provides empirical justifications for the simplifying assumptions made about our summary statistic. Sec.~\ref{sec:inference} shows, using simulated data, that our inference method performs well, especially when compared to more classical methods (Sec.~\ref{results:state_of_art}), and that it provides meaningful confidence intervals for the estimated parameter (Sec.~\ref{sec:confidence-intervals}). Sec.~\ref{seq:conclusions} presents the conclusions as well as the perspectives of this work. Fig.~\ref{fig:summary} represents graphically the structure of this article. Sec.~\ref{appendix:statistic} explains why a summary statistic more sophisticated than the one we ended up using, does not provide more information about the parameter we want to estimate. Sec.~\ref{appendix:maximum_estimation} gives all the details on our implementation of the inference part of the approach. Sec.~\ref{sec:proofs} gives explicit proofs for the few lemmas stated in the article. Sec.~\ref{sec:some-numerical-details} presents a brief overview of our software implementation and of the numerical method used, together with links to the source code.

\section{Data generation model}
\label{sec:data-generation-model}
Here we present both the modeling of a neural network as a random graph, and the simulation of the spiking activity of such a network as its stochastic evolution over time. This simulation setting yields a collection of \emph{spike trains}, which is the realization of a point process that represents the chronology of a neuron spikes. 
This simulation setting is used for two purposes :
This simulation setting is used for two purposes :
\begin{itemize}
    \item first, to simulate many data to estimate the likelihood, see Sec.~\ref{model:sbi};
    \item second, to generate additional data that will play the role of real observed data. Indeed, since our model is far from being realistic, no real data would fit into this framework.
\end{itemize}

\subsection{A Directed Erdős-Rényi model}
\label{sec:directed-ER-model}
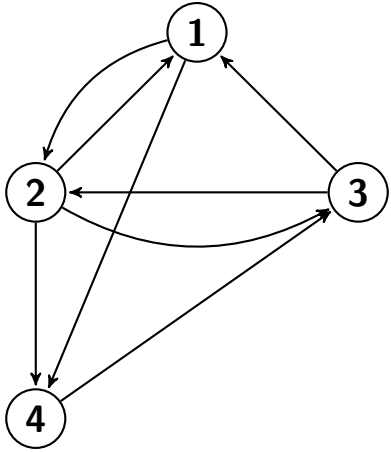
\begin{figure}[]
\begin{center}
\begin{tikzpicture}[->,>=stealth',shorten >=1pt,auto,node distance=3cm,
                    thick,main node/.style={circle,draw,font=\sffamily\Large\bfseries}]
  \node[main node] (1) {1};
  \node[main node] (2) [below left of=1] {2};
  \node[main node] (3) [below right of=1] {3};
  \node[main node] (4) [below of=2] {4};

  \path[every node/.style={font=\sffamily\small}]
    (1) edge node [left] {} (4)
        edge [bend right] node[left] {} (2)
    (2) edge node [right] {} (1)
        edge node {} (4)
        edge [bend right] node[left] {} (3)
    (3) edge node [right] {} (2)
        edge node [right] {} (1)
    (4) edge node [left] {} (3);
\end{tikzpicture}
\end{center}
\caption{Example of a directed graph representing a neural network with 4 neurons and 7 synapses.}
\label{fig:graph}
\end{figure}

The basic ingredients of our model are:
\begin{itemize}
\item a finite set $I$ of $n$ neurons, $i \in I \equiv \{1,\ldots,n\}$\footnote{We use the symbol $\equiv$ to \emph{define} its left hand side by the right hand side.},
\item a set of random \emph{synaptic weights} $W_{j \to i} \in  \{0,w\}$, for $i, j \in I$, with $w > 0$,
\item a family of  \emph{spiking probability functions} $\phi_i : \mathbb{R}  \to [0, 1 ]  , i \in I$.
\end{itemize}
Since we consider neural networks with only \emph{chemical} synapses, for which physiologists distinguish a \emph{presynaptic} and a \emph{postsynaptic} side \cite[Chap. 3]{luo:2021}, we introduce for a realization of the random graph model (denoted as $w_{i \to j}$):
\[  {\mathcal V}_{ \cdot \to i } \equiv \{ j \in I : w_{j \to i } \neq 0\} ,\]
the set of \emph{presynaptic neurons} of $i$, and
\[  {\mathcal V}_{ i \to \cdot  } \equiv \{ j \in I : w_{i \to j } \neq 0\} ,\]
the set of \emph{postsynaptic neurons} of $i$.
This defines a \emph{directed graph} in which the neurons are the vertices and the synaptic connections are the edges as illustrated on Fig.~\ref{fig:graph}. For this graph we have:
\[\mathcal{V}_{\cdot\to 1} = \{2, 3\}, \quad \mathcal{V}_{1\to\cdot} = \{2,4\}, \quad \mathcal{V}_{\cdot\to 2} = \{1, 3\}, \quad \mathcal{V}_{2\to\cdot} = \{1,3,4\}\, .\]

Our central working hypothesis is that this graph is a realization of a \emph{directed Erdős-Rényi model}, that is:

\begin{itemize}
\item $\forall i,j \in I, i \neq j, \mathbb{P}\{W_{i \to j }=w\}=p \text{ and } \mathbb{P}\{W_{i \to j }= 0\}=1-p \text{, where } p \in (0,1),$
\item the $\Big(W_{i \to j} \Big)_{\substack{i, j \in I \\ i \neq j}} \text{ are independent} $.
\end{itemize} 

\subsection{Basic discrete time model}
\label{sec:basic-discrete-time-model}
In the discrete time setting adopted in this article, our model describes the spiking activity of a finite set $I$ of neurons over time, where time is binned into small windows (to make the correspondence with real data, one can take a 1 millisecond window length). For any neuron $i \in I$, $$X_t (i ) = \left\{\begin{array}{ll}1 & \text{if neuron $i$ spikes at time $t$,} \\ 0 & \text{if neuron $i$ does not spike at time $t$.} \end{array}\right.$$ In what follows, we will simply speak of the value at time $t$ instead of speaking of the time window of index $t$.

To specify the model, we also need to introduce the notion of the \emph{last spike time of neuron i before time t}, for any $i \in I$ and $t \in \mathbb{Z}$. Formally, this is defined by 
\begin{equation}\label{eq:L_t-discrete-definition}
L_t (i) \equiv \max \{ s \le  t  : X_s (i) = 1  \} .
\end{equation}

\subsection{Membrane potential: a very simple model}
\label{sec:membrane-potential}
In this section and the next, everything is defined with respect to a given realization of the graph. When we define probabilities, we do not make this dependence on the graph realization explicit in order to keep more compact equations, but that should be clear to the reader. The \emph{membrane potential} of neuron $i$ at time $t$ is defined by: 
\begin{equation}\label{def:potential}
 V_t (i) \equiv \left\{\begin{array}{lr} \sum_{j \in {\mathcal V}_{\cdot \to i }}  w_{ j \to i} \left( \sum_{ s = L_t (i) +1 }^{ t} X_s (j)  \right)  & \text{ if  }  L_t (i ) < t, \\
0 & \text{ if  } L_t (i ) = t. \end{array}
\right.
 \end{equation}
Thus, the membrane potential value of neuron $i$ is obtained by adding up the contributions of all \emph{presynaptic neurons} $j \in {\mathcal V}_{\cdot \to i }$ \emph{of i} since its last spiking time. The membrane potential is moreover reset to $0$ at each spiking time of the neuron. Remark that with such a scheme, a neuron cannot interact with itself---more precisely there is no difference between a model where $w_{i \to i} = 0$ and a model where $w_{i \to i} \neq 0$.

\subsection{Dynamics: what makes a neuron spike?}
\label{sec:dynamics}
We start with an informal description of the dynamics assuming that we have reached time $t$:
\begin{enumerate}
\item We compute $V_t(i)$ for every neuron $i$.
\item Every neuron $i$ decides to spike at time $t+1$  with probability $\phi_i ( V_t (i ) )$, \emph{independently} of the others: \[\mathbb{P}\left\{X_{t+1} (i ) = 1 \mid \left(V_t(j)\right)_{j \in I}\right\} = \phi_i ( V_t (i ) )\,.\]
\item For every neuron $i$, we compute $V_{t+1} (i)$ according to \eqref{def:potential}.
\end{enumerate}
This algorithm can be formally translated as follows.
\begin{itemize}
\item We start at time $t=0$ from some initial condition $X_t (i ) = x_t (i )$ for all $t \le 0, i \in I$.
\item We suppose that for all $i \in I$, there exists $l_i \le 0$,  such that $x_{l_i} (i ) = 1$. This means  that $l_i \le L_0 (i) \le 0$ is well-defined for any $i \in I$, and that we are able to compute $V_0 ( i)$ for each neuron $i$. We call such a past configuration an \emph{admissible past}.
\item We consider a family of uniform random variables $U_t (i ) , i \in I, t  \geq 1$, which are  independent and identically distributed (IID), with a uniform distribution on $[0, 1 ]$ ($U_t(i) \sim \mathcal{U}(0,1)$).
\item Then we define in a recursive way for every $t \geq 0$:
  \begin{equation}\label{def:fixedpast}
  X_{t+1 } (i )=\left\lbrace
		\begin{array}{ll}
			  1  & \mbox{if } U_{t+1}  ( i)  \le \phi_i (V_{t }(i ) ),  \\
      	 0  & \mbox{if } U_{t+1} ( i)  >  \phi_i ( V_{t }(i ) ),
		\end{array}
	        \right.
  \end{equation}
  where for each $t\geq 1$ and $i\in I$, $V_{t}( i )$ is the membrane potential of neuron $i$ at the previous time step  $t$, defined according to (\ref{def:potential}).
\end{itemize}
It is easy to show \cite[Chap. 2]{galves.locherbach.pouzat:2024} that the process $(\mathbb{V}_t)_{t \geq 0 }$, $\mathbb{V}_t  =( V_t (i),i \in I)$, is a Markov chain on $\mathbb{R}^n$ and is therefore more suitable for simulations. We can see $\mathbb{V}_{t}$ as a vector valued random variable:
\begin{equation}\label{eq:bigV}
\mathbb{V}_{t} \equiv \begin{bmatrix}
V_t(1) \\
V_t(2) \\
\vdots \\
V_t(n)
\end{bmatrix}.
\end{equation}
The transitions of the Markov chain  $(\mathbb{V}_{t})_{t} \geq 0$ can be described as follows:
 \begin{equation}\label{eq:simple-discrete-V-evolution}
V_{t+1 } (i )=\left\lbrace
		\begin{array}{ll}
			  0 & \mbox{if } U_{t+1} ( i)  \le \phi_i (V_{t }(i ) ),  \\
      	 V_{t} (i ) + \sum_{j \in {\mathcal V}_{\cdot \to i } } w_{ j\to i } \mathbb{1}_{ \{ U_{t+1} ( j)  \le \phi_j (V_{t }(j ))  \}}& \mbox{if } U_{t+1} ( i)  >  \phi_i ( V_{t }(i ) ) 	.
		\end{array}
	\right.
\end{equation}
Both equations above (Eq.~\ref{def:fixedpast} and Eq.~\ref{eq:simple-discrete-V-evolution}) can be put together into :
\begin{equation}\label{eq:explicit_v}V_{t+1} (i ) = ( 1 - X_{t+1}(i) ) \left[ V_t (i ) + \sum_{j \in {\mathcal V}_{\cdot \to i } } w_{j \to i } X_{t+1} (j ) \right].\end{equation}

\begin{figure}[]%
\centering
\includegraphics[scale=0.75]{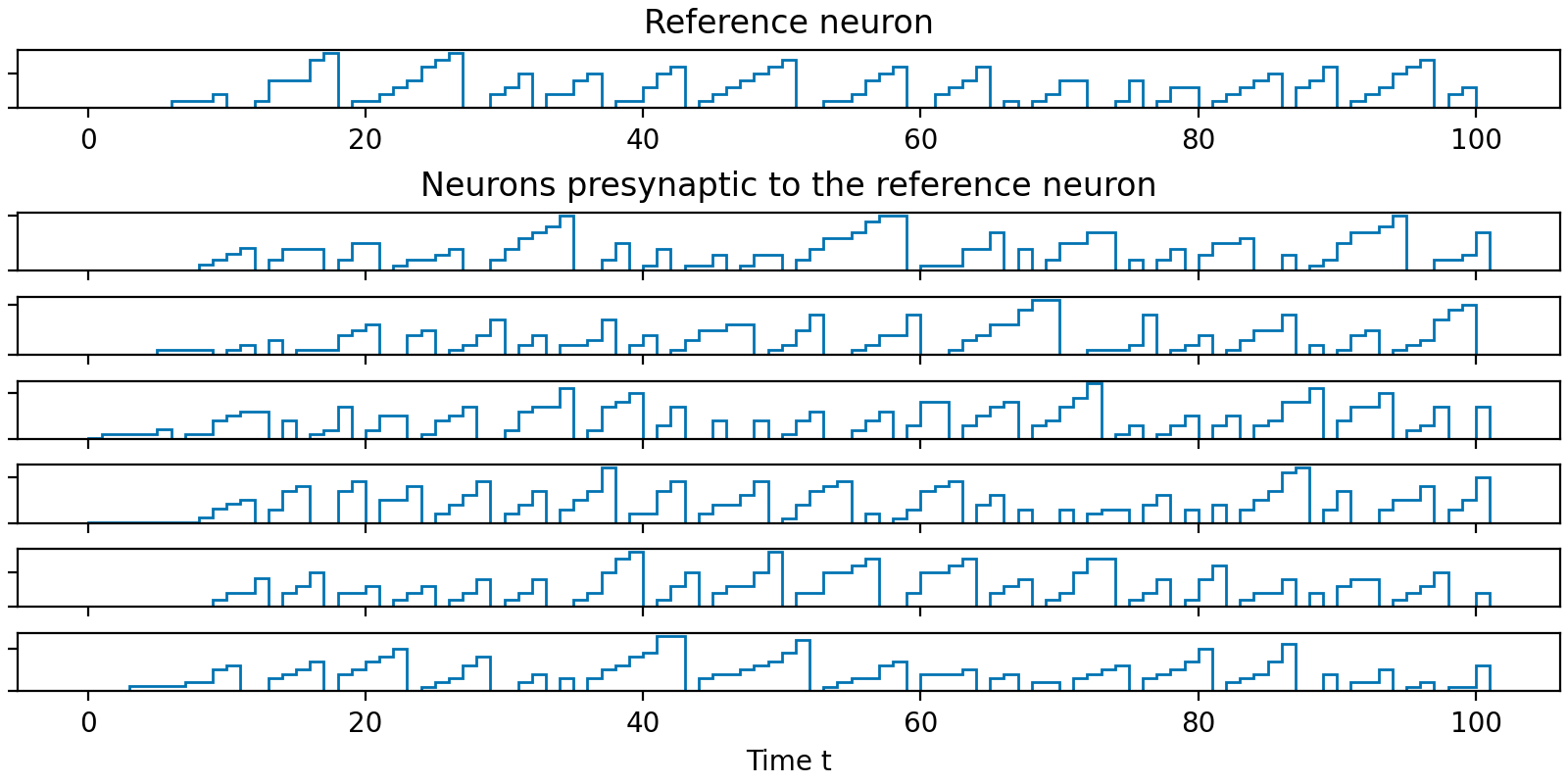}
\caption{Trajectories of the membrane potential of a reference neuron and its presynaptic partners.}
\label{fig:membrane_potential_evol}
\end{figure}

Fig.~\ref{fig:membrane_potential_evol} shows an example of simulated trajectories of the membrane potentials of a neuron and of its presynaptic partners. 

The spiking behaviour of the neuron $i$ is then completely determined by the values of $X_t(i)$ for all $t$. The previously mentionned associated point process, the realization of which is a spike train is \begin{equation}\label{eq:spike_train}Y_i \equiv \{t \in \mathbb{N} \, | \, X_t(i) = 1 \}.\end{equation}

\subsection{Simple setting considered in this article}
\label{sec:simple-setting-considered-in-this-article}

In the sequel, the following model parameters are set to the following values :
\begin{itemize}
\item Number of neurons: $n=1000$.
\item Synaptic weight: $w = 0.01$.
\item Spiking probability functions: $ \forall i, \phi_i = \phi$, where $\phi(x) \equiv \max\left(0,\min(x,1)\right)$.
\end{itemize}
\emph{The only parameter to be estimated from the data is then the connection probability $p$}. Notice that with the directed Erdős-Rényi graph model (Sec.~\ref{sec:directed-ER-model}) combined with the Galves-Löcherbach model (Eq.~\ref{def:fixedpast} and \ref{eq:simple-discrete-V-evolution}) and a positive synaptic weight $w$, an increase in $p$ will translate into typically more presynaptic partners for a given neuron, leading to an increased activity (more spikes) of the latter. 

\section{Parameter estimation with a Monte Carlo method}
\label{sec:general_method}
The typical setting considered in this article is a network of $n=1000$ neurons, from which 10 neurons are ``randomly selected'' and recorded for $T=10^6$ time steps (if we consider that our discretization time window is 1~ms long, this number of steps corresponds to roughly 15 minutes in real time; that is not long compared to the actual recordings that typically last hours). The actual network we have in mind for latter application is the \emph{antennal lobe} (the first olfactory relay) of the locust, \emph{Schistocerca americana} \cite{Laurent_1996}. Roughly 1100 neurons are found in this network. 

\subsection{An intractable likelihood}
\label{sec:an-intractable-likelihood}

 The spike train $Y_i$ of a neuron $i$, described in Eq.~\ref{eq:spike_train}, is a random variable in the set $\mathcal{S} \equiv \mathcal{T}^*$, where $\mathcal{T} \equiv \{1, \dots, T\}$. We write as $\mathbf{Y} \equiv \left(Y_{i_1}, \ldots, Y_{i_s}\right)$ the random variable on $\mathcal{S}^s$ which consists of the spike trains of $s$ different randomly chosen neurons $\{i_1,\ldots,i_s\} \subset I$, which play the role of real data.
Since the neurons are randomly chosen, the $Y_j$ are clearly identically distributed, and we will sometimes refer to $Y$ as a random variable following this distribution (different from $\mathbf{Y}$ which is a random variable following the joint distribution).

Given a realization of $\mathbf{Y}$, \textit{i.e.} a sample $\mathbf{y} = \left(y_1, \ldots, y_s\right) \in \mathcal{S}^s$ of spike trains corresponding to a sample of observed neurons, our goal is to estimate the value of $p$, the connection probability between neurons in the whole neural network the observations stem from.

Notice that the spike trains are not independent, since the neurons are part of the same neural network, and are directly or indirectly connected with each other. There is moreover a dependency between the observed spike trains, $\mathbf{Y}$, and the unobserved ones, $\mathbf{Y}^u \equiv \left(Y_i\right)_{i \in I \setminus \{i_1,\ldots,i_s\}}$. The probability distribution of $\mathbf{Y}$:
\begin{equation}
\label{eq:likelihood_untractable}
\mathbb{P}_p\{\mathbf{Y}=\mathbf{y}\} = \sum_{\mathbf{y}^u} \mathbb{P}_p\{\mathbf{Y}=\mathbf{y},\mathbf{Y}^u=\mathbf{y}^u\}\end{equation}
is therefore not tractable. The classical maximum likelihood approach consists of viewing $\mathbb{P}_p\{\mathbf{Y}=\mathbf{y}\}$ as a function of $p$ with $\mathbf{y}$ already observed (and therefore fixed) \cite[Def. 6.3.1]{casella.berger:2002}; and in choosing as an estimate of $p$ \cite[Def. 7.2.4]{casella.berger:2002}:\[\argmax_{p\in (0,1)} \mathbb{P}_p\{\mathbf{Y}=\mathbf{y}\}.\] In views of Eq.~\ref{eq:likelihood_untractable}, this approach cannot be implemented as such.

\subsection{Using the sampling distribution of a ``well chosen'' statistic in place of the likelihood}
\label{general_method:transformation}

Our intractability problem is fortunately not new, it is almost systematically met by practitioners of \emph{Approximate Bayesian Computation} (ABC) \cite{Sisson_2018} and, more generally, by \emph{Simulation-Based Inference} (SBI)\footnote{ABC can be considered as a specific instance of SBI.} users \cite{Cranmer_2020}. This means that working solutions, as opposed to optimal ones, have been proposed. This is the path we are going to follow here.

We look for a summary statistic $T(\mathbf{Y})$ such that, the distribution $\mathbb{P}_p\left\{T(\mathbf{Y})\right\}$ (the \emph{sampling distribution} of $T(\mathbf{Y})$ \cite[Def. 5.2.1]{casella.berger:2002}) is ``sensitive'' to the value of $p$, with a ``small'' variance (see Fig.~\ref{fig:hist_alpha} and Fig.~\ref{fig:hist_spikefreq} in appendix). Assuming we can compute or estimate this sampling distribution, we will solve:
\[\argmax_{p\in (0,1)} \mathbb{P}_p\left\{T(\mathbf{Y})=T(\mathbf{y})\right\} \quad \text{   instead of   } \quad \argmax_{p\in (0,1)} \mathbb{P}_p\{\mathbf{Y}=\mathbf{y}\}\, .\]

Notice that if $T(\mathbf{Y})$ is sufficient for $p$ \cite[Def. 6.2.1 and Theo. 6.2.2]{casella.berger:2002}, given an observation $\mathbf{y}$, the two alternatives will lead to the same location of the maximum (that is, the same estimated value for $p$). Since we cannot even get $\mathbb{P}_p\{ \mathbf{Y}\}$, we cannot hope finding a sufficient statistic.  We will rely instead on a statistic $T(\mathbf{Y})$ found ``by hand'', inspired by the statistics commonly used by neurophysiologists (Sec.~\ref{appendix:statistic}). We will then empirically check that $\argmax_{p\in (0,1)} \mathbb{P}_p\left\{T(\mathbf{Y})=T(\mathbf{y})\right\}$ is close to the target $p$ and exhibits suitable statistical properties.

\subsection{Particular statistics considered in this article}
\label{general_method:particular_case}

Let us consider the particular case where the statistic $T(\mathbf{Y})$ is of the form \[T(\mathbf{Y}) = T\left(Y_1, \ldots, Y_s\right) = \left(\varphi\left(Y_1\right), \ldots, \varphi\left(Y_s\right)\right),\] for some real-valued function $\varphi : \mathcal{S} \to \mathbb{R}$.  
A first example for this function $\varphi$ is the empirical frequency of spikes per time step, \textit{i.e.} the number of spikes of the neuron divided by the total number of steps of the record. An other example is the estimated shape parameter (or alpha parameter) of a gamma density fitted to the inter-spike interval (ISI) distribution  (see Sec.~\ref{statistic:gamma_param} for definition). The choice of these statistics is detailed in Sec.~\ref{appendix:statistic}. 

Even if the $Y_1, \ldots, Y_s$ are sampled from the same network time evolution, we treat the transformed variables $\varphi\left(Y_1\right), \ldots, \varphi\left(Y_s\right)$ as IID variables, with sampling distribution $\mathbb{P}_p\left\{\varphi(Y)=\varphi(y)\right\}$ (see Sec.~\ref{ssec:estimating_likelihood:likelihood} for an empirical justification). 

The function $p \mapsto \mathbb{P}_p\left\{T(\mathbf{Y})=T(\mathbf{y})\right\}$ is then approximated by a \emph{composite marginal likelihood} \cite{varin.reid.firth:2011}, that is the product of the sampling distributions of the $\varphi(Y_j), \, j=1, \dots, s$ when the model parameter is $p$. It can be expressed as:
\begin{equation}\label{eq:loglikelihood_indep}
     \mathbb{P}_p\left\{T(\mathbf{Y})=T(\mathbf{y})\right\} = \mathbb{P}_p\left\{\left(\varphi\left(Y_1\right), \ldots, \varphi\left(Y_s\right)\right) = \left(\varphi\left(y_1\right), \ldots, \varphi\left(y_s\right)\right)\right\} = \prod_{j=1}^s \mathbb{P}_p\left\{\varphi(Y_j)=\varphi(y_j)\right\}.
   \end{equation}

   We have therefore moved our problem from solving:
   \[\argmax_{p\in (0,1)} \mathbb{P}_p\{\mathbf{Y}=\mathbf{y}\}\quad \text{ to the simpler one of solving } \quad   \argmax_{p\in (0,1)} \prod_{j=1}^s \mathbb{P}_p\left\{\varphi(Y_j)=\varphi(y_j)\right\}\, .\]

\subsection{Simulation-based inference}
\label{model:sbi}
The ``sad'' reality is that we still cannot obtain an explicit expression of $\mathbb{P}_p\left\{\varphi(Y)=\varphi(y)\right\}$, but our model specification can be viewed as a recipe for simulation. We can therefore use simulations to get an estimate $\widehat{\mathbb{P}}_p\left\{\varphi(Y)=\varphi(y)\right\}$ of $\mathbb{P}_p\left\{\varphi(Y)=\varphi(y)\right\}$.
We proceed as follows:
\begin{itemize}
\item The interval of physiologically relevant $p$ values is discretized such that\\ $p \in \mathcal{P} \equiv \left\{0.005 + 0.001\times i, \; i=0,1,\ldots, 95 \right\}$\footnote{This choice of grid is made to avoid too small values of $p$, for which a lot of neurons tends to be isolated (${\mathcal V}_{ \cdot \to i } = \emptyset$) and spike at most only once.}. 
\item For each connection probability $p$ in $\mathcal{P}$, a number of $K=50$ independent directed graphs are generated, each with $n = 1000$ neurons.
\item For each graph, the time evolution is simulated during $T = 10^6$ time steps, with an initial membrane potential value of $v_0=0.01$.
  \begin{itemize}
  \item This yields $K \times n = 50000$ spike trains, on which the $\varphi$ function is applied.
  \item This gives a set $\left\{ \varphi\left(y_i^{(k)}\right) : \ 1\leq i \leq n,\ 1 \leq k \leq K\right\}$ of values that we treat as $n\times K$ IID samples from the same distribution, \textit{i.e.} the distribution of the random variable $\varphi(Y)$.
  \item If the distribution of the $\left\{ \varphi\left(y_i^{(k)}\right) : \ 1\leq i \leq n,\ 1 \leq k \leq K\right\}$ is clearly non-Gaussian, the histogram is used as a nonparametric estimator of $\mathbb{P}_p\left\{\varphi(Y)=\varphi(y)\right\}$, this is (a frequentist version of) the \emph{classical ABC method}; otherwise a Gaussian distribution is fitted and a parametric estimator of the sampling distribution is used, yielding a \emph{synthetic likelihood} \cite{Wood_2010, frazier:2023}.
  \end{itemize}
\item This procedure results in a set of estimated sampling distributions: $\left\{\widehat{\mathbb{P}}_p\left\{\varphi(Y)=\varphi(y)\right\}\right\}_{p \in \mathcal{P}}$.
\end{itemize}

The inference performances are checked by choosing a $p$ in the range of $\mathcal{P}$, \emph{but not necessarily in that set} (see Sec.~\ref{sec:inference}). A directed Erdős-Rényi graph is generated and its dynamics is simulated exactly as described above. A predefined number of 5, 10, 15 or 20 spike trains are selected at random and their statistic $\varphi(y)$ are computed. The maximum location in $\mathcal{P}$, denoted as $\tilde{p}$ is defined as: \[\tilde{p} \equiv \argmax_{p\in \mathcal{P}} \prod_{j=1}^s \widehat{\mathbb{P}}_p\left\{\varphi(Y_j)=\varphi(y_j)\right\}\; \text{ with } \; s=5,10,15,20\, .\] The estimate $\hat{p}$ of $p$ is then defined as the location of the maximum of the quadratic polynomial interpolating $\prod_{j=1}^s \widehat{\mathbb{P}}_p\left\{\varphi(Y_j)=\varphi(y_j)\right\}$ between $\tilde{p}$ and its two nearest neighbors (see Sec.~\ref{appendix:maximum_estimation} for details).  

\section{Estimating \texorpdfstring{$\mathbb{P}_p\left\{\varphi(Y)=\varphi(y)\right\}$}{the likelihood}}
\label{sec:estimating_likelihood}
In this section, we start by empirically verifying the assumption made in Sec.~\ref{general_method:particular_case}, then provide evidence that we can estimate $\prod_{j=1}^s \mathbb{P}_p\left\{\varphi(Y_j)=\varphi(y_j)\right\}$. 

\subsection{Justification for the independence of the \texorpdfstring{$\varphi(Y_i)$}{summary statistic on the spike trains}}
\label{ssec:estimating_likelihood:indep}
In Sec.~\ref{general_method:particular_case}, in order to simplify both the function to estimate, and the practical estimation of this function, we made the assumption that, when sampling $s$ random spike trains $Y_1,\dots, Y_s$ from the time evolution of our model, the random variables $\varphi(Y_1), \dots, \varphi(Y_s)$ are independent. The random variables $Y_1, \dots, Y_n$ are clearly not independent, but we make the assumption that the dependence structure is largely lost under the projection $\varphi$ on the real line, making the transformed variables $\varphi(Y_1), \dots, \varphi(Y_s)$ approximately independent. While full mutual independence cannot be empirically verified, we provide here empirical evidence suggesting that treating these variables as independent is a reasonable approximation.

\begin{enumerate}
\item \textbf{The variables are pairwise approximately uncorrelated.} We show this by randomly sampling $N = 10000$ times a couple of spike trains $(Y_i, Y_j)$ such that $Y_i$ and $Y_j$ are from the same time evolution of the same (random) graph, and estimate the correlation coefficient of the couple $\left(\varphi(Y_i), \varphi(Y_j)\right)$. Fig.~\ref{fig:pairwise_corrcoef} shows that this correlation coefficient is very close to zero. We also show that, even in the worst case, \textit{i.e.} when $Y_j$ is randomly selected among the post synaptic neurons of $Y_i$ (we identify the spike trains with the corresponding neuron here), the correlation coefficient stay relatively small, especially for large values of $p$.
    
\begin{figure}[]
\centering
\begin{subfigure}{.49\textwidth}
  \centering
  \includegraphics[width=1\linewidth]{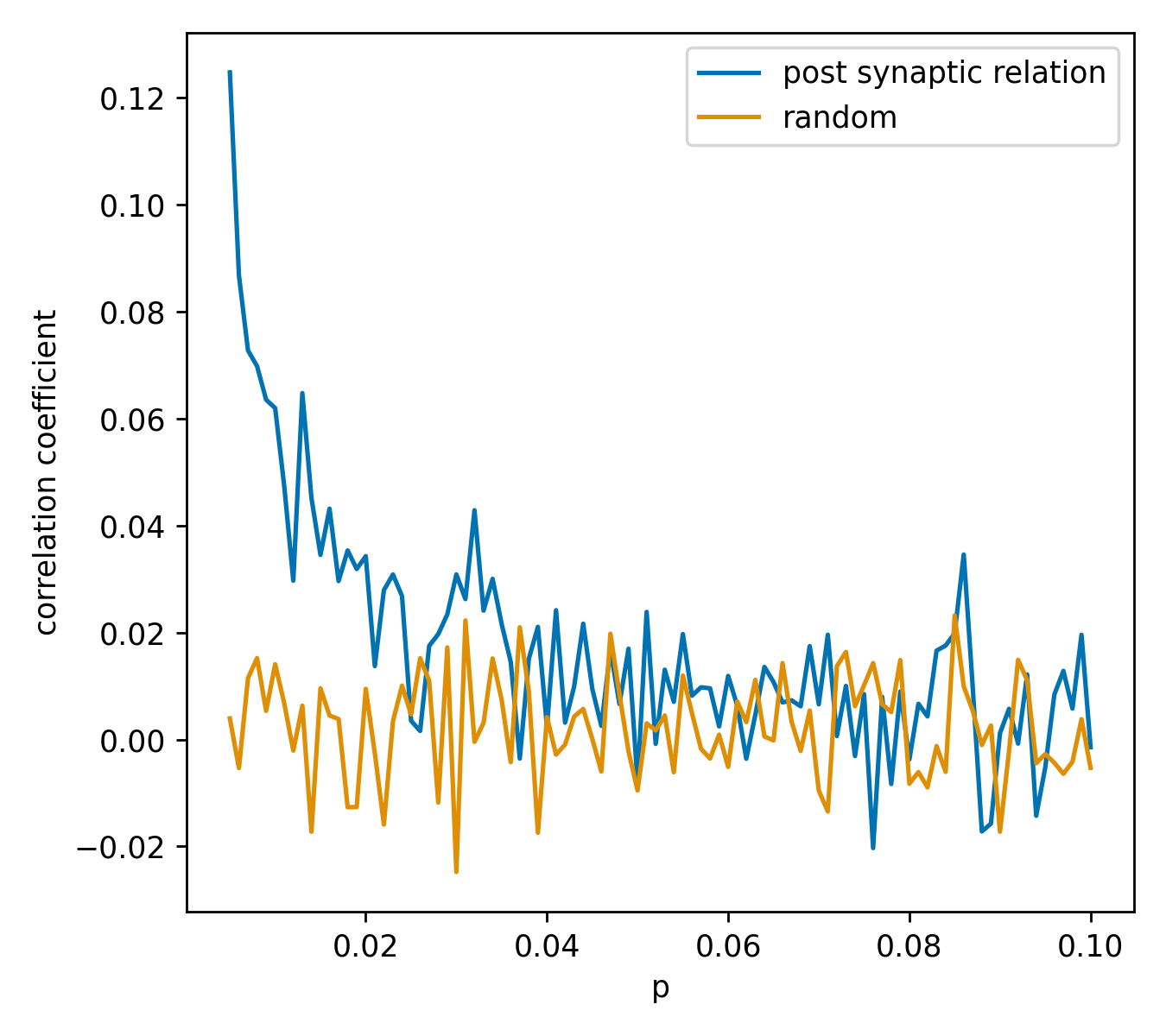}
  \caption{$\varphi$ is the frequency of spikes}
  \label{fig:pairwise_corrcoef:spikefreq}
\end{subfigure}
\begin{subfigure}{.49\textwidth}
  \centering
  \includegraphics[width=1\linewidth]{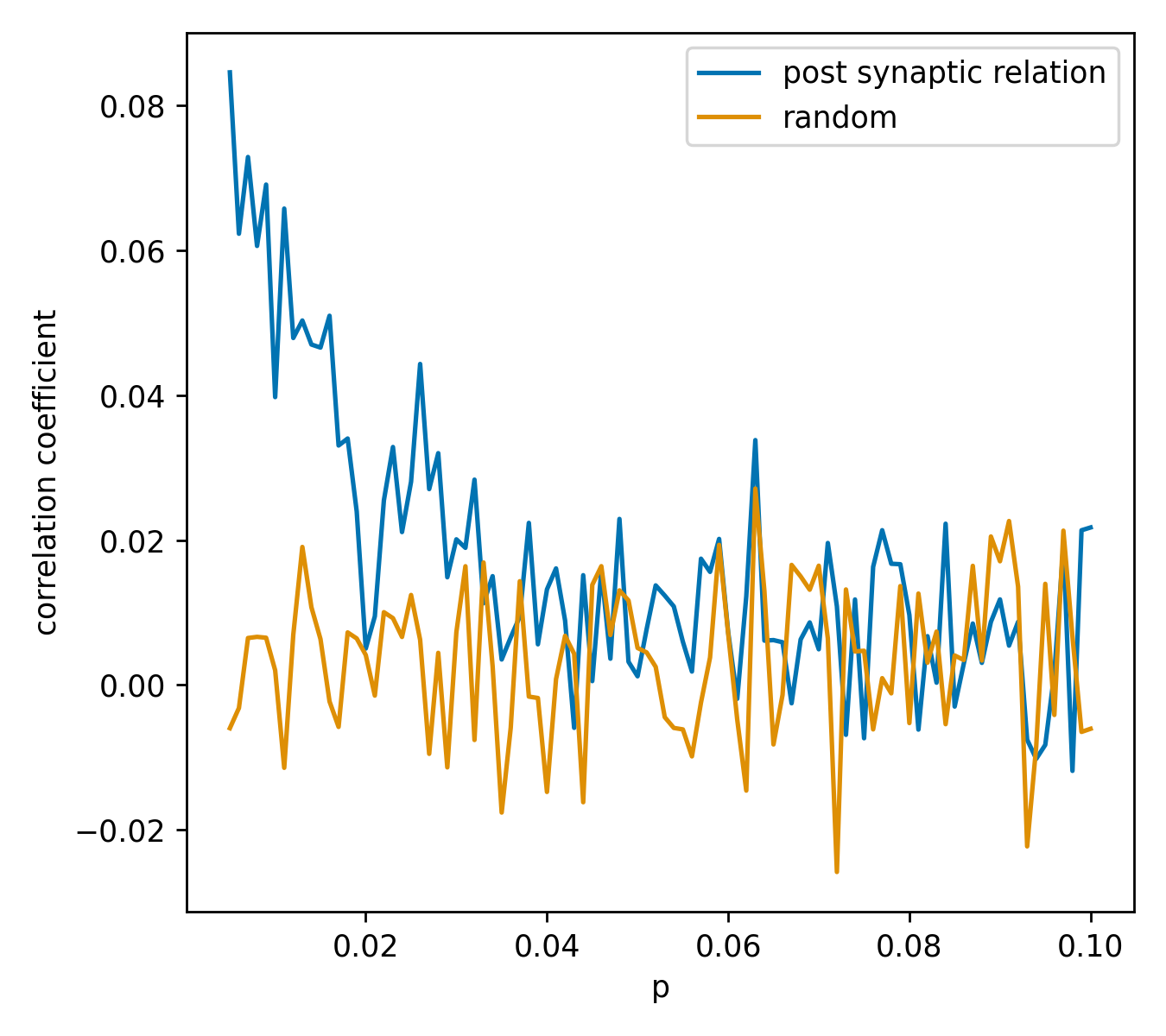}
  \caption{$\varphi$ is the estimated $\alpha$ parameter}
  \label{fig:pairwise_corrcoef:alphas}
\end{subfigure}
\caption{Estimated correlation coefficient between $\varphi(Y_i)$ and $\varphi(Y_j)$, when the couple $(Y_i, Y_j) $ is randomly sampled (\texttt{random} label in the legend) or when it is sampled such that $Y_i$ is presynaptic to $Y_j$ (\texttt{post\_synaptic\_relation} label in the legend). }
\label{fig:pairwise_corrcoef}
\end{figure}

\item \textbf{The variables are pairwise approximately jointly Gaussian.} By plotting the same random samples of the random variable $\left(\varphi(Y_i), \varphi(Y_j)\right)$ as in point 1, we see that their joint distribution is very close to a joint Gaussian distribution. Fig.~\ref{fig:jointplot_post_synaptic_selection} shows an example of this for $p = 0.012$. It is still a good approximation if the couple is sampled such that $Y_i$ is presynaptic to $Y_j$.

\begin{figure}[]
\centering
\begin{subfigure}{.49\textwidth}
  \centering
  \includegraphics[width=1\linewidth]{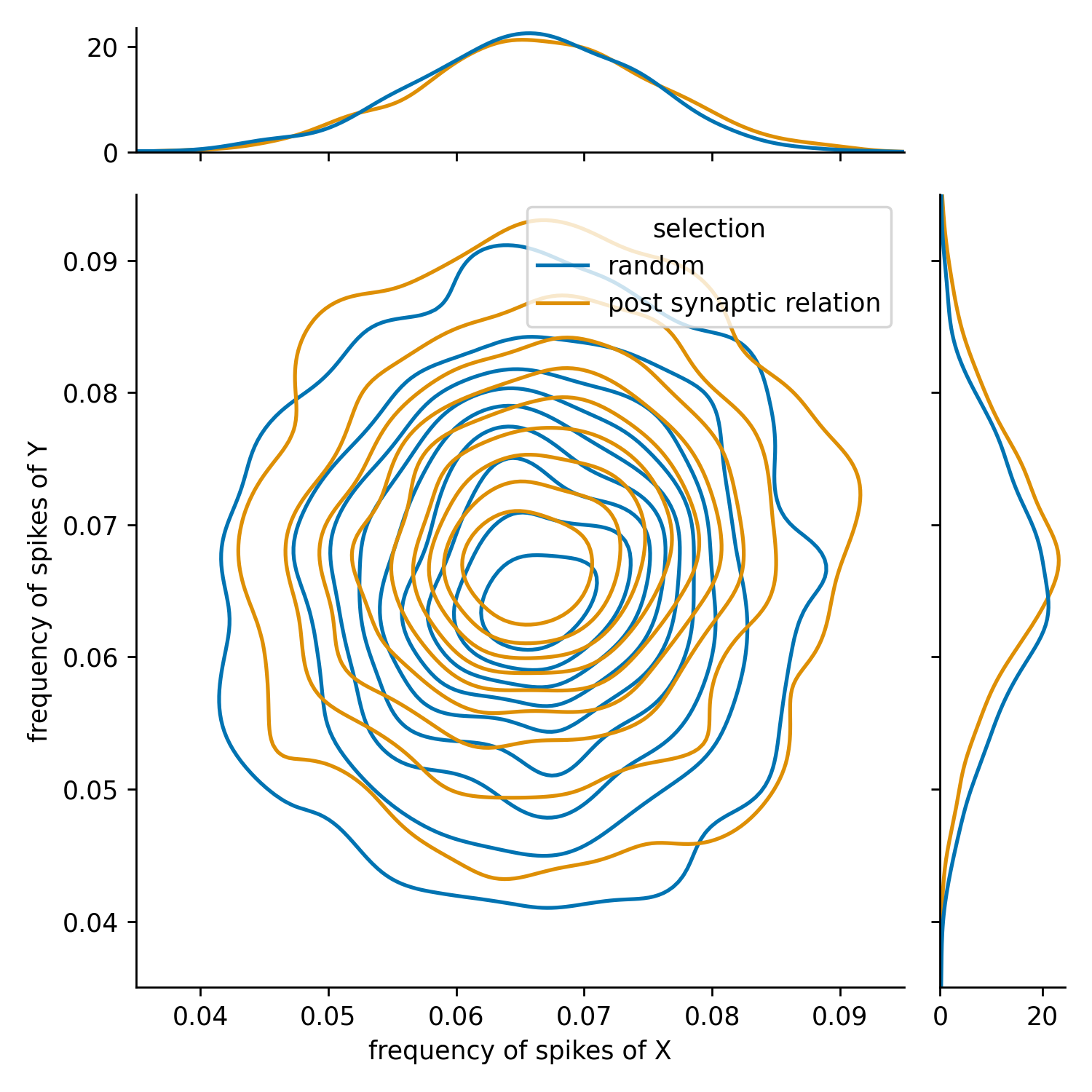}
  \caption{$\varphi$ is the frequency of spikes}
  \label{fig:jointplot_post_synaptic_selection:spikefreq}
\end{subfigure}
\begin{subfigure}{.49\textwidth}
  \centering
  \includegraphics[width=1\linewidth]{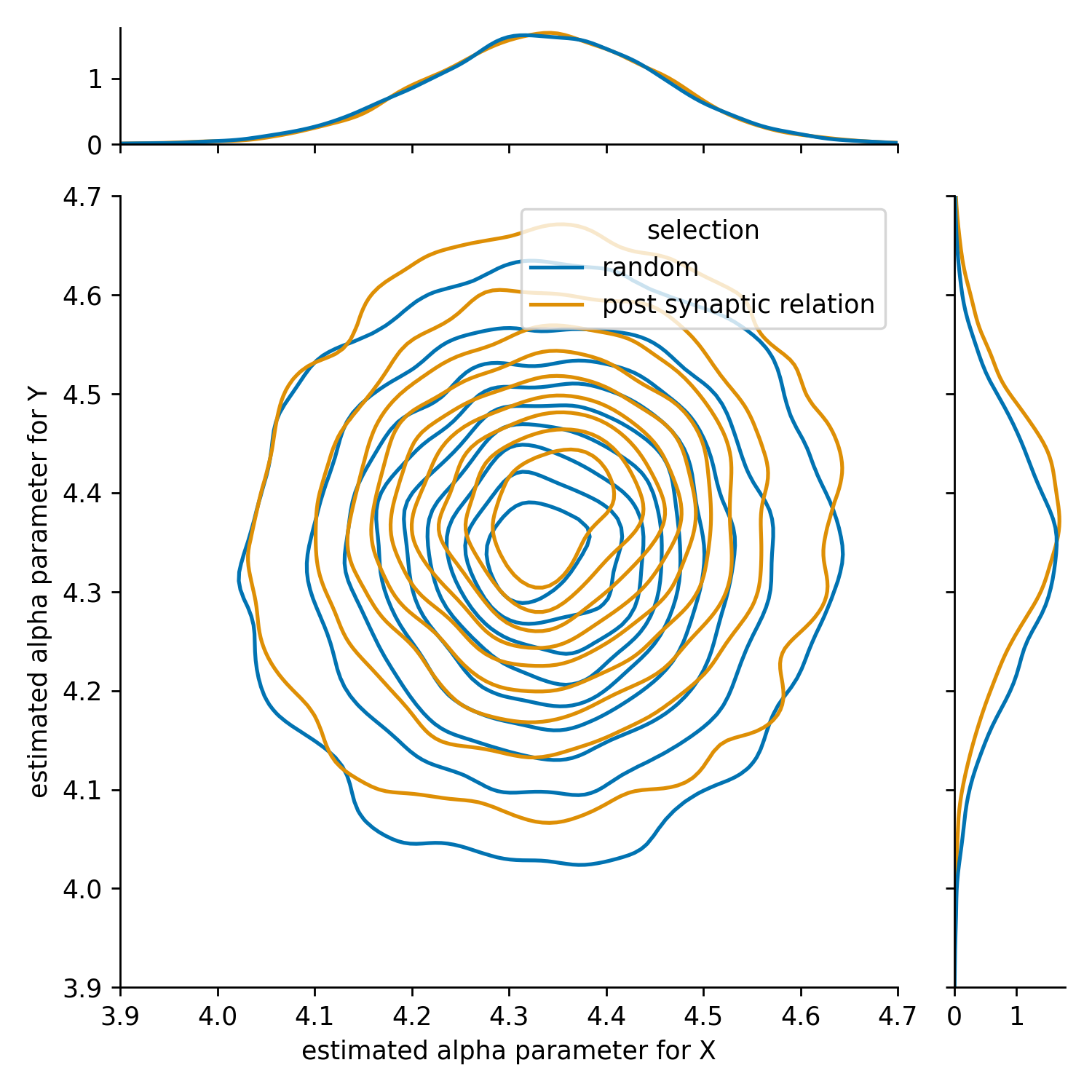}
  \caption{$\varphi$ is the estimated $\alpha$ parameter}
  \label{fig:jointplot_post_synaptic_selection:alphas}
\end{subfigure}
\caption{Estimated joint and marginal distributions of (a) the spike frequency and (b) the estimated alpha parameter of the ISI distributions of the summary statistic of two randomly selected neurons $Y_i$ and $Y_j$, from the same time evolution of the same neural network with $p = 0.012$ (\texttt{random} label in the legend). Also shows the distribution when the two neurons are sampled such that $Y_i$ is presynaptic to $Y_j$ (\texttt{post\_synaptic\_relation} label in the legend)}
\label{fig:jointplot_post_synaptic_selection}
\end{figure}

\item \textbf{The variables are approximately jointly Gaussian.} This is a stronger claim than the previous one. If the random variables $\varphi(Y_1), \dots, \varphi(Y_s)$ are jointly Gaussian, then the square of their Mahalanobis distance \cite[p. 21]{Ripley_1996} to their mean should follow the chi-squared distribution with $s$ degrees of freedom. Fig.~\ref{fig:chisquared_compare} shows that it is approximately true for a large range of $p$ values.

\begin{figure}[]
\centering
\begin{subfigure}{.49\textwidth}
  \centering
  \includegraphics[width=1\linewidth]{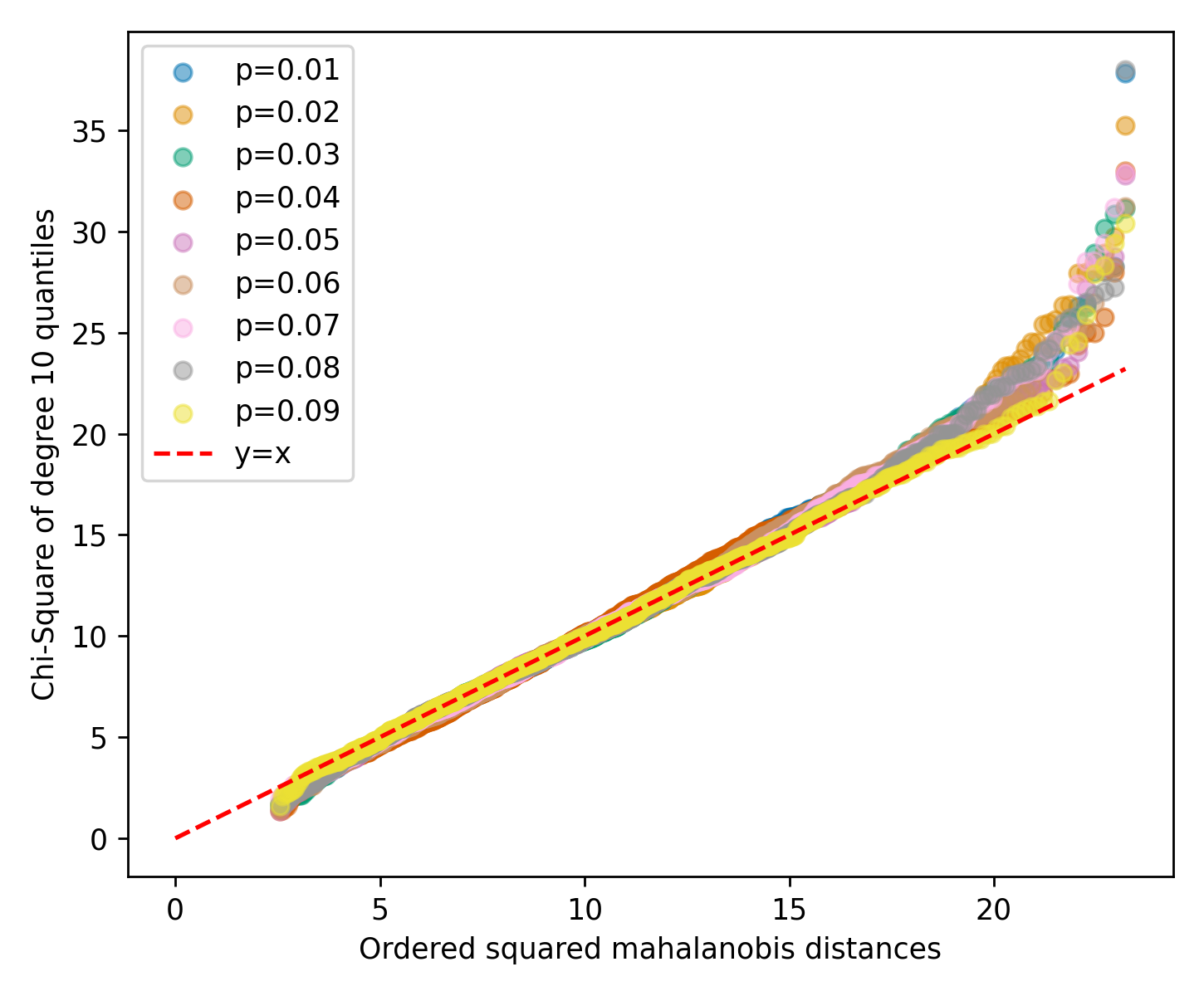}
  \caption{$\varphi$ is the frequency of spikes}
  \label{fig:chisquared_compare:spikefreq}
\end{subfigure}
\begin{subfigure}{.49\textwidth}
  \centering
  \includegraphics[width=1\linewidth]{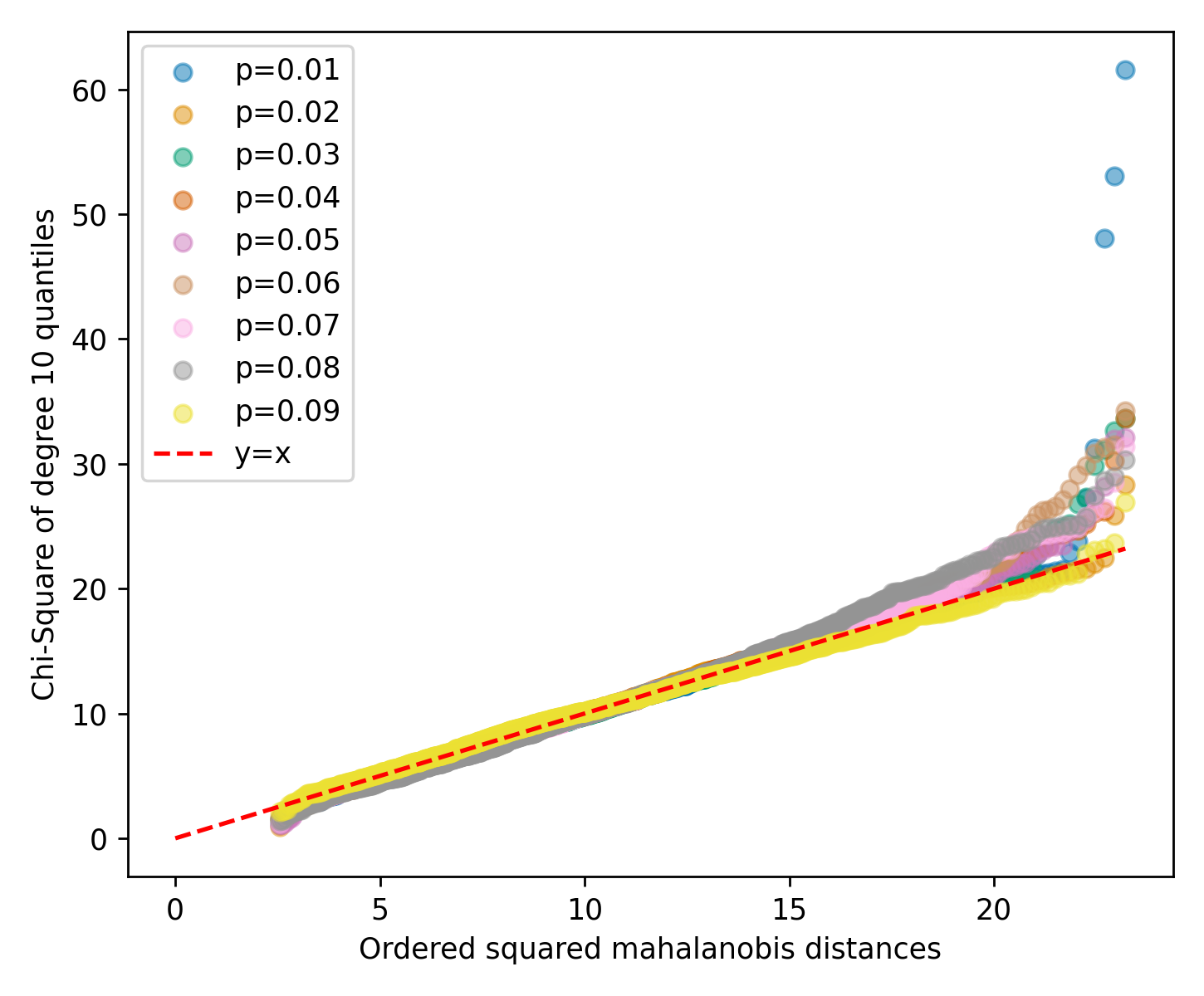}
  \caption{$\varphi$ is the estimated $\alpha$ parameter}
  \label{fig:chisquared_compare:alphas}
\end{subfigure}
\caption{Q-Q plot between the squared Mahalanobis distance of $\varphi(Y_1), \dots, \varphi(Y_s)$ and chi-squared distribution with $s$ degrees of freedom, with $s = 10$.}
\label{fig:chisquared_compare}
\end{figure}

\end{enumerate}

These three points suggest that the random variables $\varphi(Y_1), \dots, \varphi(Y_s)$ are nearly pairwise uncorrelated and jointly Gaussian, thus, approximately independent. Since both the maximum likelihood method, and the kernel density estimator are robust to small dependencies in the data \cite{TRAN1990193,bhat}, these variables can be treated as independent for the estimations. Equation \ref{eq:loglikelihood_indep} is then approximately true, and estimating $\mathbb{P}_p\left\{T(\mathbf{y})\right\}$ reduces to estimating $\prod_{j=1}^s \mathbb{P}_p\left\{\varphi(Y_j)=\varphi(y_j)\right\}$.

\begin{remark}
   The fact that the summary statistics are approximately independent suggests that the system is close to a mean-field limit \cite{fournier.locherbach:2016, Galves2020}. This means that each neuron has approximately the same behaviour, independently from the other neurons, which is the behaviour of a ''typical neuron'' in a hypothetical infinite network.
\end{remark}

\subsection{Estimation of \texorpdfstring{$\mathbb{P}_p\left\{\varphi(Y)=\varphi(y)\right\}$}{the likelihood of the summary statistics on the spike trains}}
\label{ssec:estimating_likelihood:likelihood}

\begin{figure}[]
\centering
\begin{subfigure}{.49\textwidth}
  \centering
  \includegraphics[width=1\linewidth]{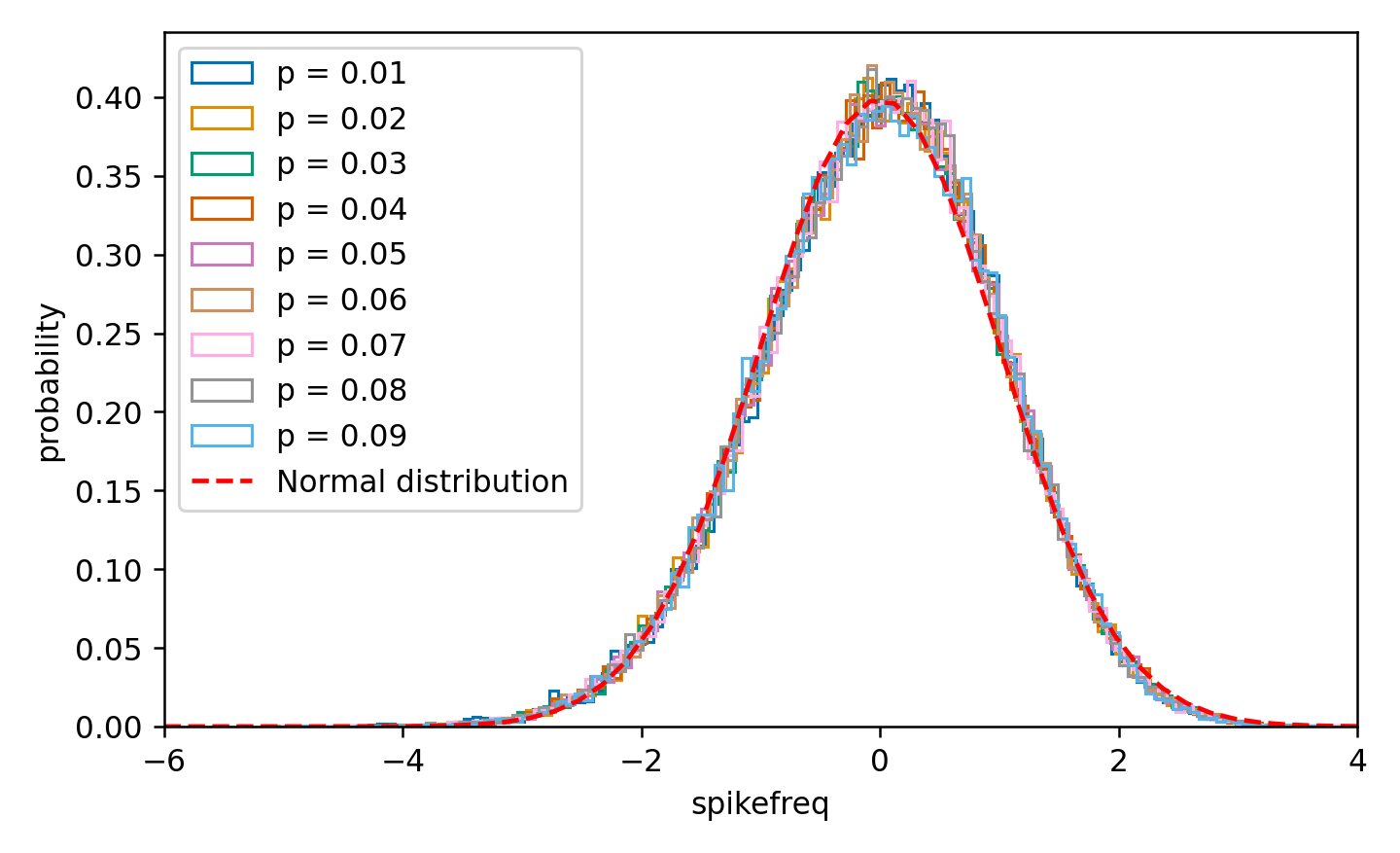}
  \caption{$\varphi$ is the frequency of spikes}
  \label{fig:gaussian:spikefreq_cumulative}
\end{subfigure}
\begin{subfigure}{.49\textwidth}
  \centering
  \includegraphics[width=1\linewidth]{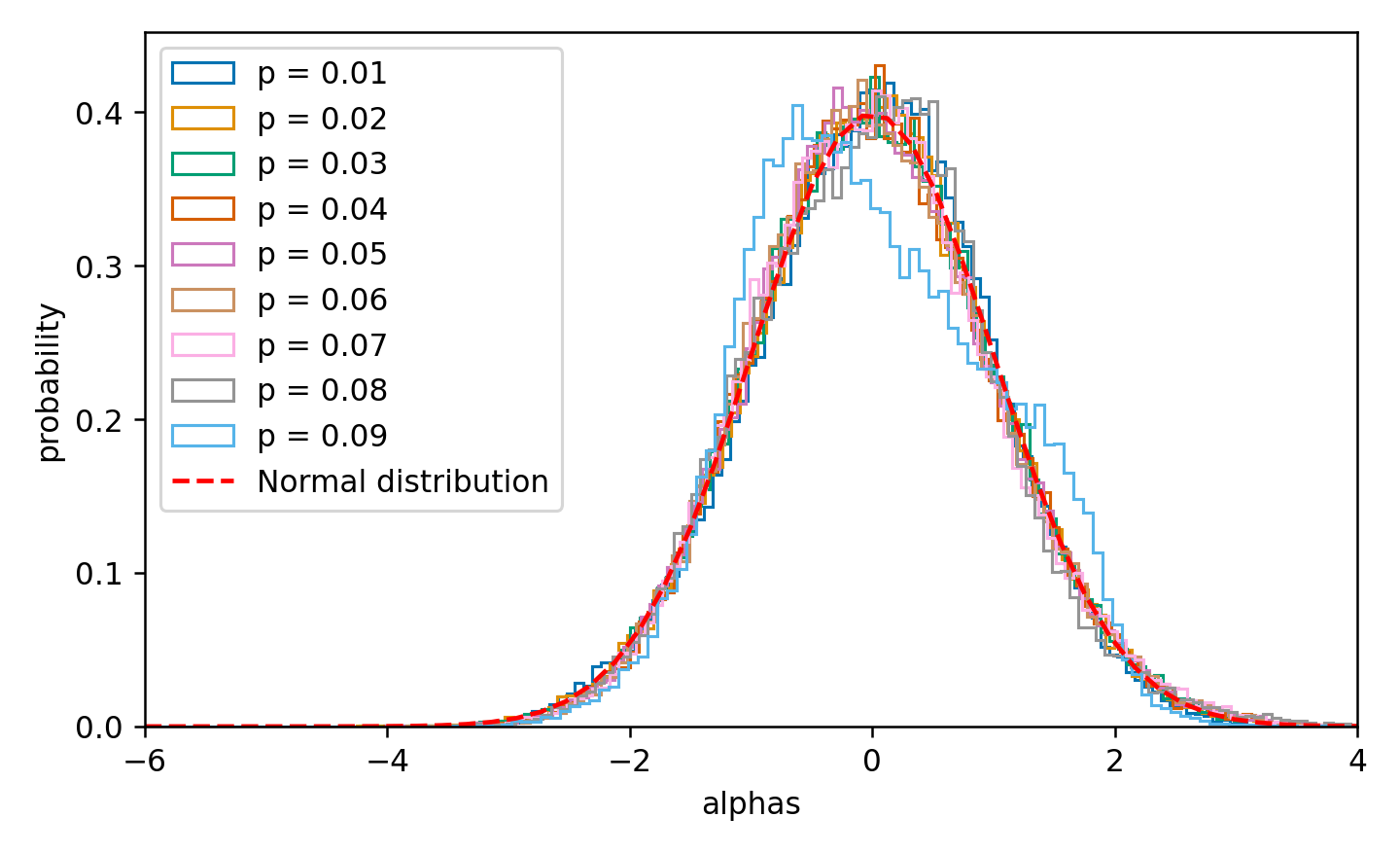}
  \caption{$\varphi$ is the estimated $\alpha$ parameter}
  \label{fig:gaussian:alphas_cumulative}
\end{subfigure}
\caption{Sampling distributions for different values of $p$}
\label{fig:gaussian}
\end{figure}

\begin{figure}[]
\centering
\begin{subfigure}{.49\textwidth}
  \centering
  \includegraphics[width=1\linewidth]{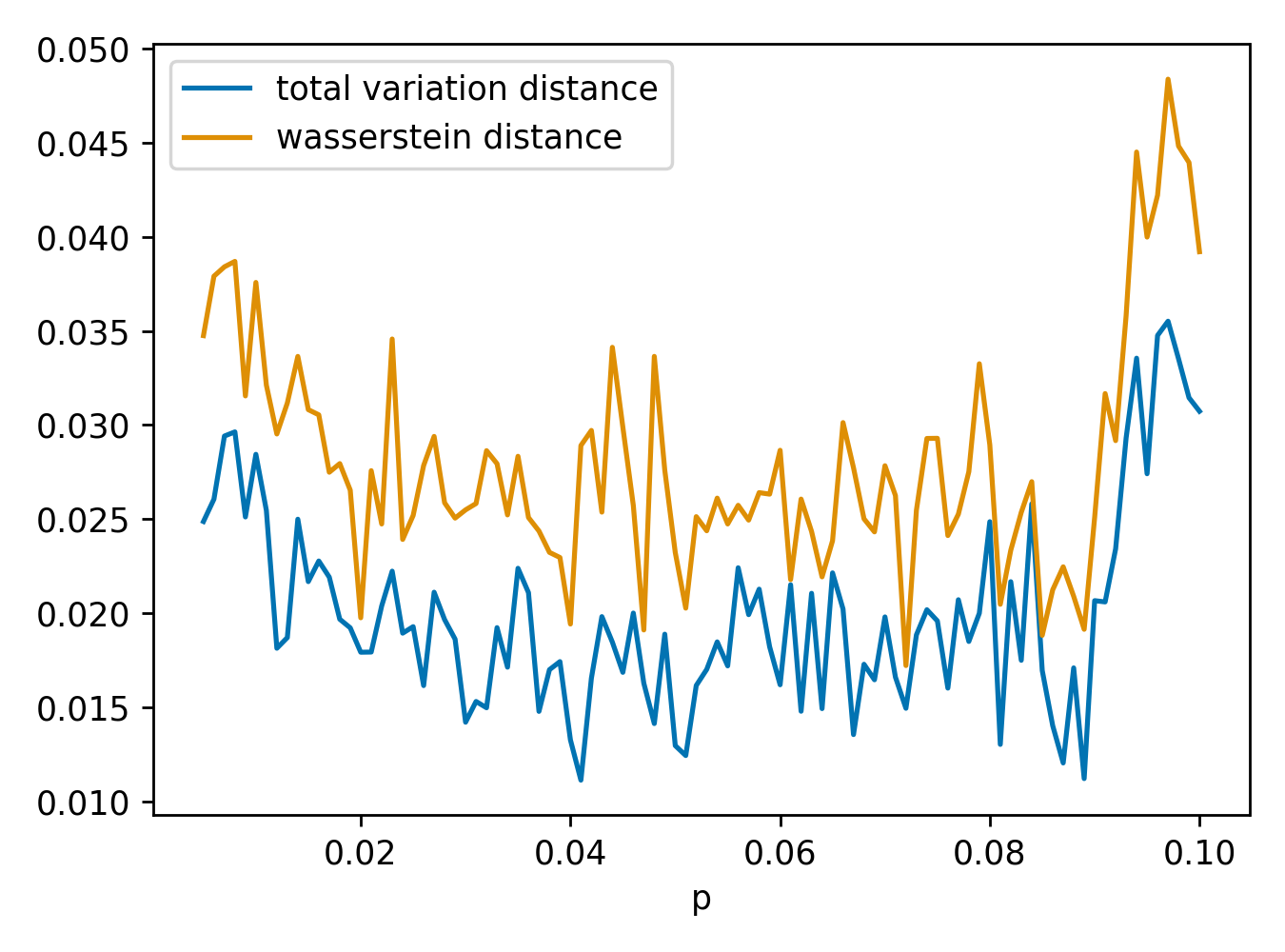}
  \caption{$\varphi$ is the frequency of spikes.}
  \label{fig:dist_to_gaussian:spikefreq}
\end{subfigure}
\begin{subfigure}{.49\textwidth}
  \centering
  \includegraphics[width=1\linewidth]{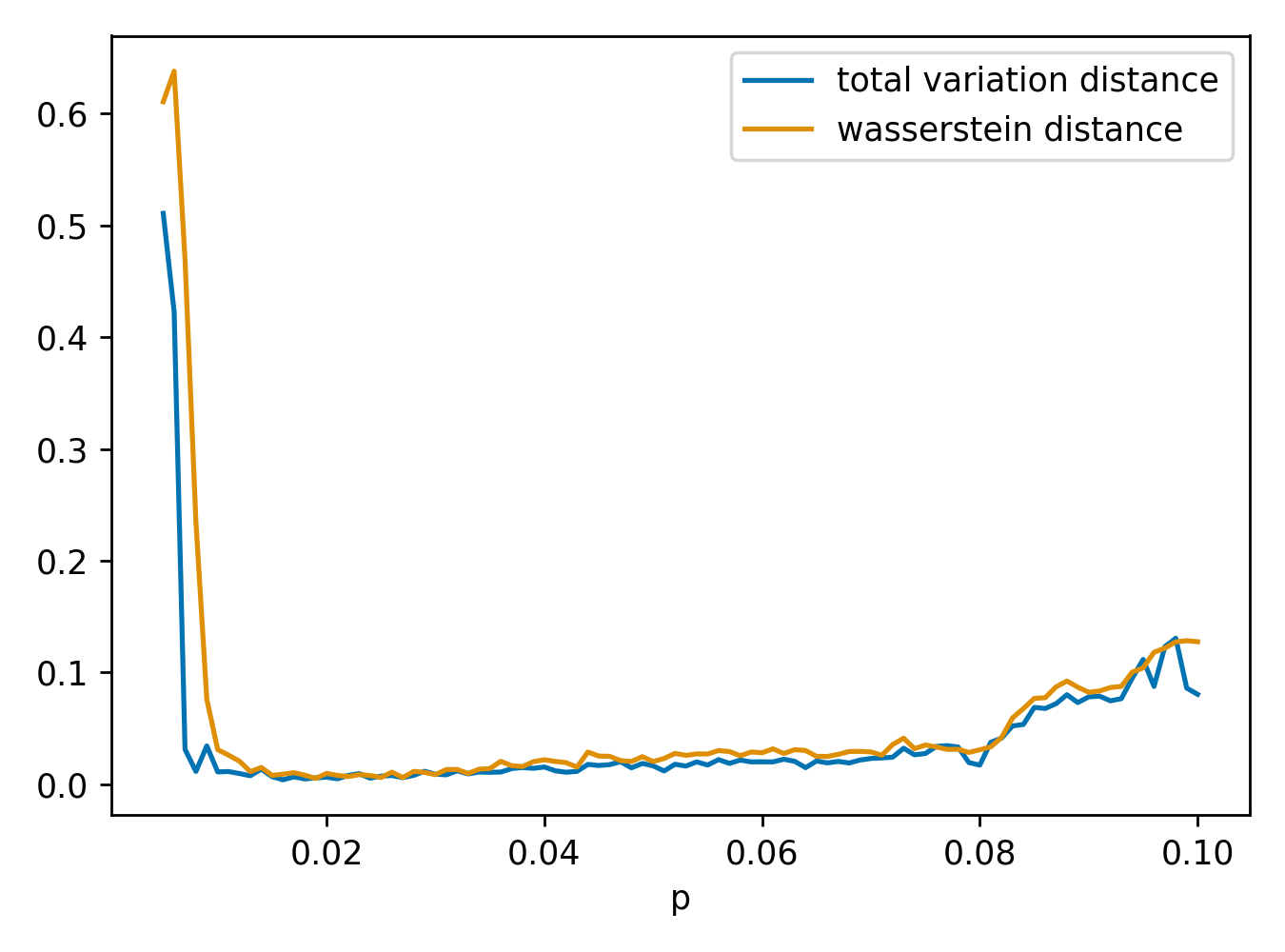}
  \caption{$\varphi$ is the estimated $\alpha$ parameter}
  \label{fig:dist_to_gaussian:alphas}
\end{subfigure}
\caption{Total variation and Wasserstein distance between the distribution of $\varphi(\mathbf{y})$ and the Gaussian distribution, as functions of $p$.}
\label{fig:dist_to_gaussian}
\end{figure}

Results from Sec.~\ref{ssec:estimating_likelihood:indep} suggest that this distribution is a Gaussian distribution. Using the samples described above, we confirm this by comparing the probability density function (Fig.~\ref{fig:gaussian}) and by computing the distance between the actual distribution and a Gaussian distribution with estimated parameters (Fig.~\ref{fig:dist_to_gaussian}). Thus, the parametric estimation of $\mathbb{P}_p\left\{\varphi(Y)=\varphi(y)\right\}$ only consists of the estimation of the mean $\mu_p$ and the variance $\sigma_p^2$ of this Gaussian distribution.

However, as we can see in Fig.~\ref{fig:dist_to_gaussian}, if $\varphi$ is the estimated alpha parameter of the ISI distribution, for either small or large values of $p$, the distance between the actual distribution of $\varphi(Y)$ and a Gaussian distribution can be quite large. This results in larger error in the estimation (see Sec.~\ref{sec:inference}), in which case estimating $\mathbb{P}_p\left\{\varphi(Y)=\varphi(y)\right\}$ with a nonparametric estimator (a histogram) is a satisfactory alternative.

\section{Inference of parameter \texorpdfstring{$p$}{p}}
\label{sec:inference}
\begin{figure}[]
\centering
\begin{subfigure}{.49\textwidth}
  \centering
  \includegraphics[width=1\linewidth]{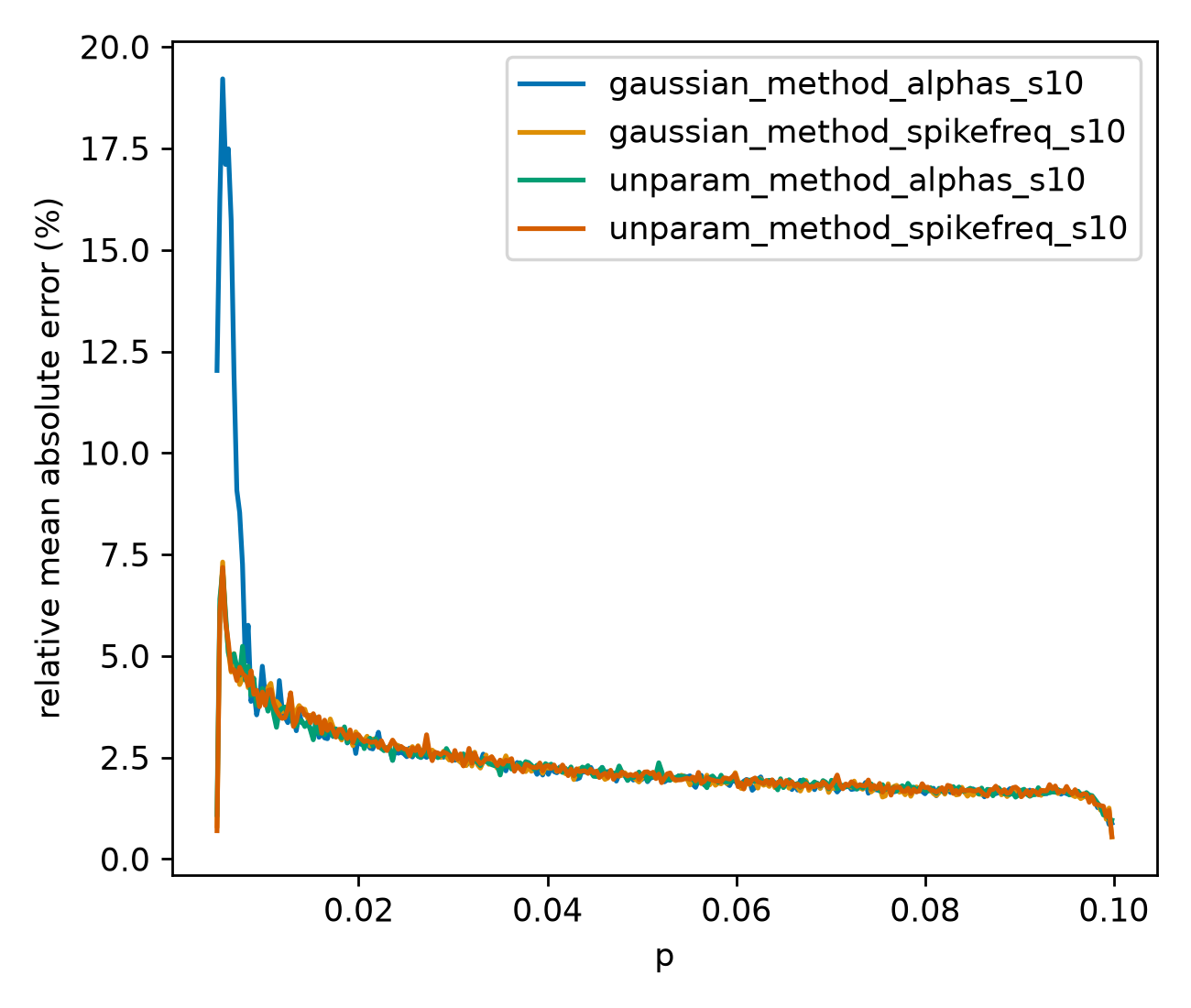}
  \caption{Mean absolute relative error}
  \label{fig:compare_methods:mae}
\end{subfigure}
\begin{subfigure}{.49\textwidth}
  \centering
  \includegraphics[width=1\linewidth]{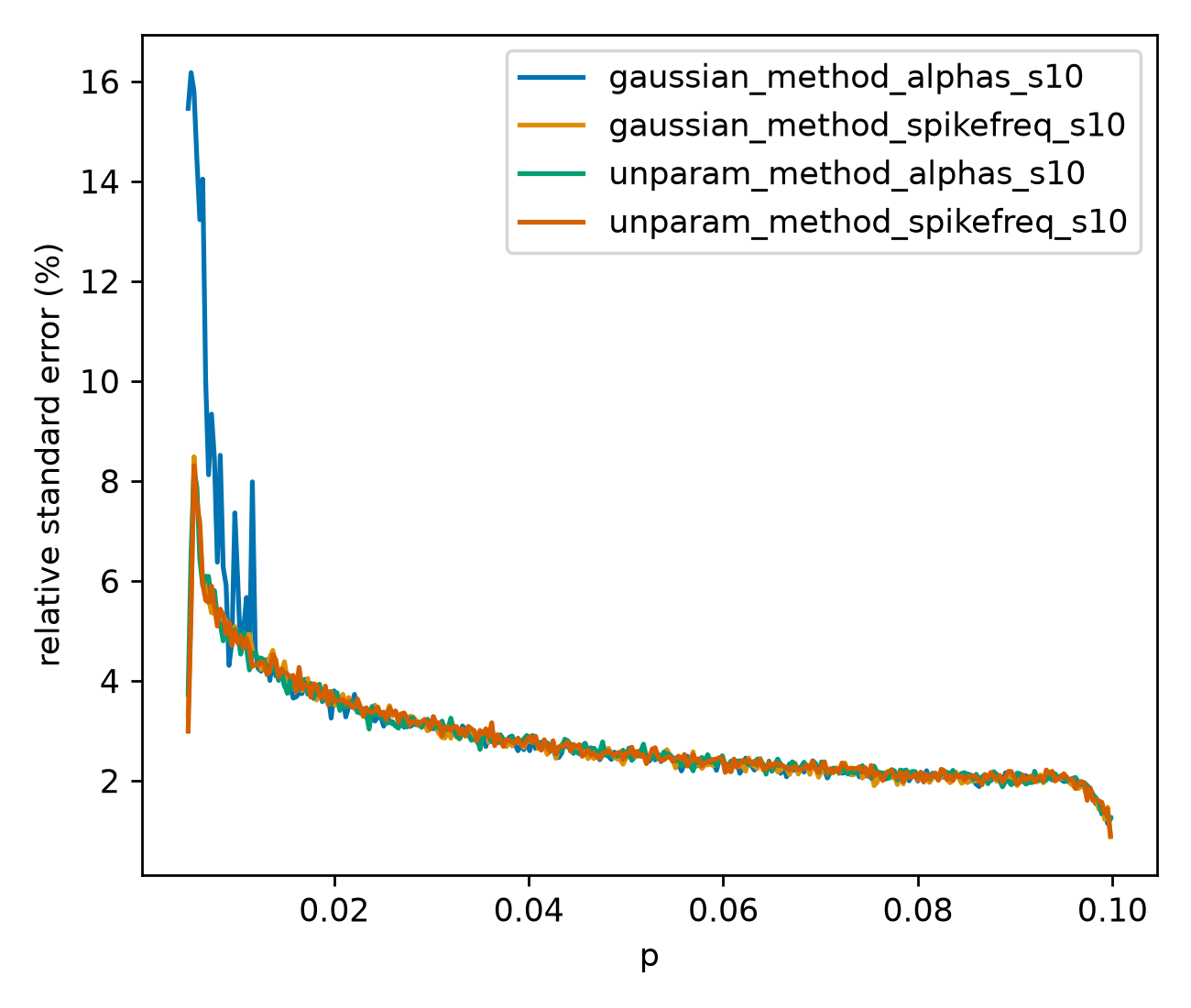}
  \caption{Relative standard error}
  \label{fig:compare_methods:std}
\end{subfigure}
\caption{Relative mean absolute error (a) and standard error (b) of the estimation of $p$ for variants of our method, with $s=10$.}
\label{fig:compare_methods}
\end{figure}

Here, we present the results of our method. We used the two different $\varphi$ functions discussed before, and presented in Sec.~\ref{appendix:statistic}, the frequency of spikes, labeled as \texttt{spikefreq}, and the shape parameter of the ISI distribution, labelled as \texttt{alpha}. The estimation of $\mathbb{P}_p\left\{\varphi(Y)=\varphi(y)\right\}$ is either parametric (Gaussian), labeled as \texttt{gaussian\_method}, or nonparametric, with the computation of a histogram, labelled as \texttt{unparam\_method}. The first result (Fig.~\ref{fig:compare_methods}) is presented with this four variants, but after this we focus on the \texttt{gaussian\_method\_spikefreq} variant (and sometimes the  \texttt{unparam\_method\_alpha} variant), for readibility.

The experiments are conducted as follows. An independent neural network with $n = 1000$ neurons and a given connection probability $p$ in $\mathcal{P}' = \big\{0.005 + 0.0003\cdot i, \quad 0 \leq i \leq 316 \big\}$\footnote{The grid on which the method is evaluated was chosen to be denser for two reasons : the "actual values" of $p$ for which the method is evaluated are different than the values of $p$ used for model characterization; less simulations were required for each value of $p$.} is generated and its time evolution is simulated for $T = 10^6$ time steps.
A number $N_e = 1000$ of estimations of $p$ is done, using $N_e$ randomly chosen samples of $s=10$ spike trains, from the corresponding network. This is repeated independently for each value of $p$ in $\mathcal{P}'$.

Fig.~\ref{fig:compare_methods} shows the mean absolute relative error (in percentage), with respect to $p$. 
We see that these variants of our method give very similar results, except for the \texttt{alphas\_gaussian\_likelihood method}, for small values of $p$.
Indeed, for small values of $p$, the distribution of $\varphi(Y)$ is clearly not Gaussian, as we can see in Fig.~\ref{fig:dist_to_gaussian:alphas}.  

\begin{figure}[]
\centering
\begin{subfigure}{.49\textwidth}
  \centering
  \includegraphics[width=1\linewidth]{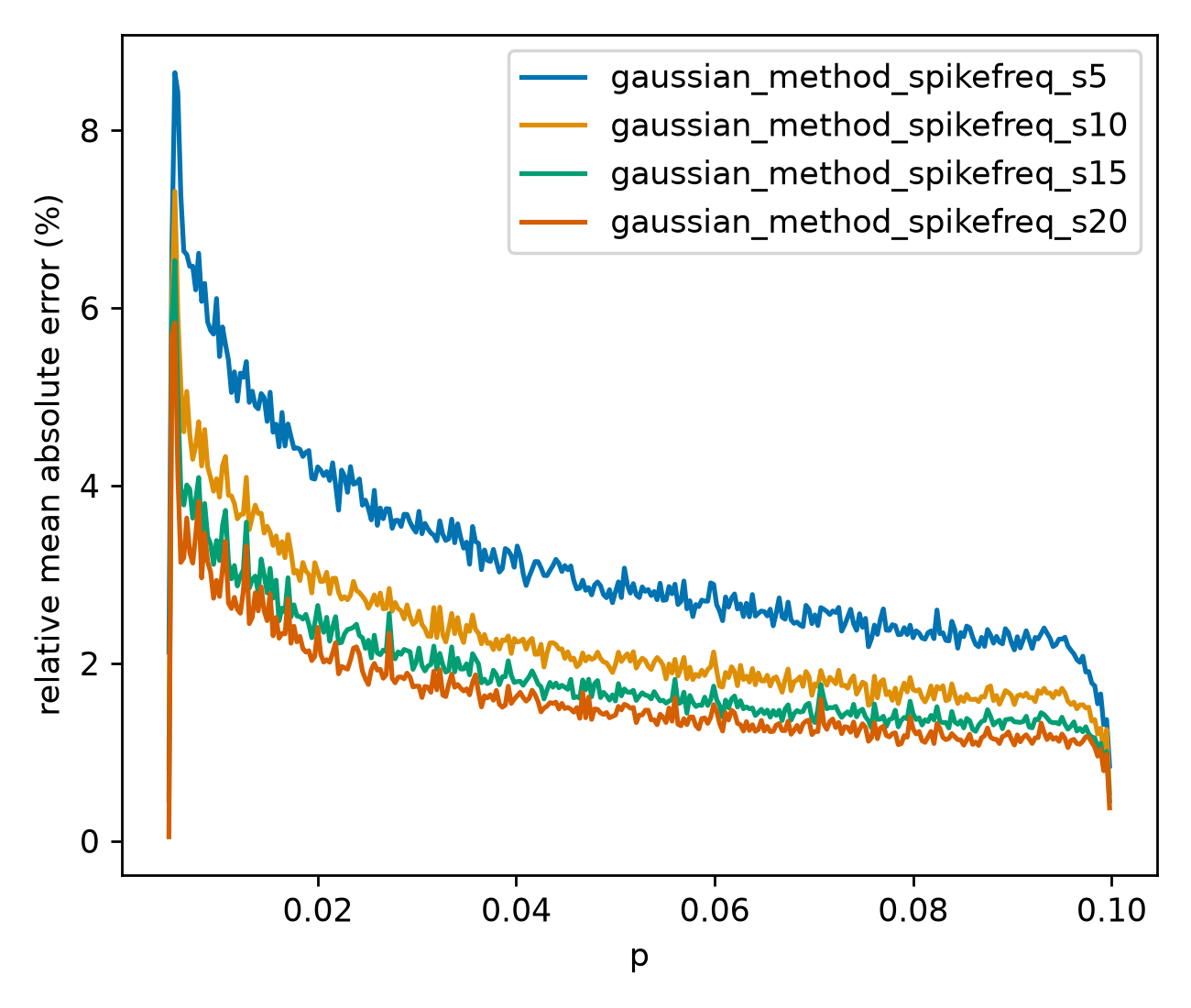}
  \caption{Mean absolute relative error}
  \label{fig:compare_sample_size:mae}
\end{subfigure}
\begin{subfigure}{.49\textwidth}
  \centering
  \includegraphics[width=1\linewidth]{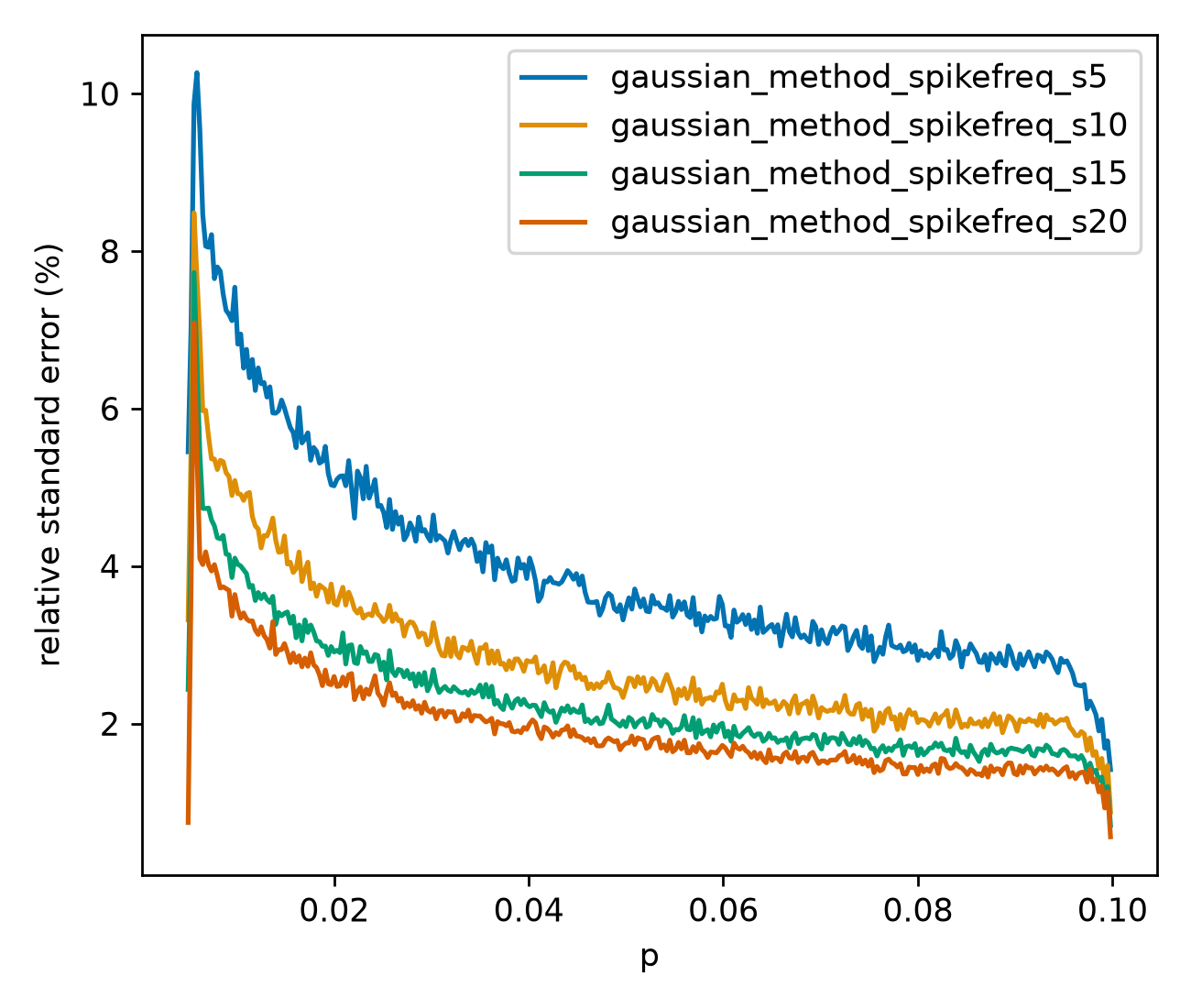} 
  \caption{Relative standard error}
  \label{fig:compare_sample_size:std}
\end{subfigure}
\caption{Relative mean absolute error (a) and standard error (b) of the estimation of $p$ for our method with $\varphi$ being the frequency of spikes and with parametric Gaussian likelihood, for different values of the sample size $s$.}
\label{fig:compare_sample_size}
\end{figure}

We also compared the performance of the method for different sample sizes ranging from $s=5$ to $s=20$, results are shown in Fig.~\ref{fig:compare_sample_size}.

\subsection{Comparison with the classical method}
\label{results:state_of_art}
The classical method for estimating the connection proportion $p$ consists of reconstructing the subgraph formed by the observed neurons. The subgraph reconstruction is usually done by computing a cross-correlation of their spike trains, and by deciding that there is a connection if this cross-correlation is above some threshold \cite{Bartho_2004}, it can also be done with a more sophisticated, model-dependent, approach \cite{De_Santis_2022} and \cite[Chap. 7]{galves.locherbach.pouzat:2024}. Given this estimation for each couple of neurons in the sample, we denote $\widehat{N_c}$ the estimation of $N_c$, the number of connections between neuron pairs in the sample. The estimation of $p$ is then simply: \[\widehat{p} = \frac{\widehat{N_c}}{s(s-1)}\,.\] We call this kind of method \emph{graph reconstruction method}. 

\begin{lemma}
\label{lemma:binomial}
Let $s\in \mathbb{N}$ with $s\geq 2$ and $p\in (0,1)$.
In the optimal case, if the realization $N_c$ were to be known, in which case we take $\widehat{N_c} = N_c$, then the optimal mean absolute error of $\widehat{p}$ is $$\mathbb{E}\big[|\widehat{p} - p|\big] = 2\sum_{k=0}^{\lceil s(s-1)p\rceil-1} \binom{s(s-1)}{k} p^k(1-p)^{s(s-1) - k}\Big(p-\frac{k}{s(s-1)}\Big).$$ 
The optimal standard error is $$\sigma(\widehat{p}) = \sqrt{\frac{p(1-p)}{s(s-1)}}.$$
\end{lemma}

\begin{figure}[]
\centering
\begin{subfigure}{.49\textwidth}
  \centering
  \includegraphics[width=1\linewidth]{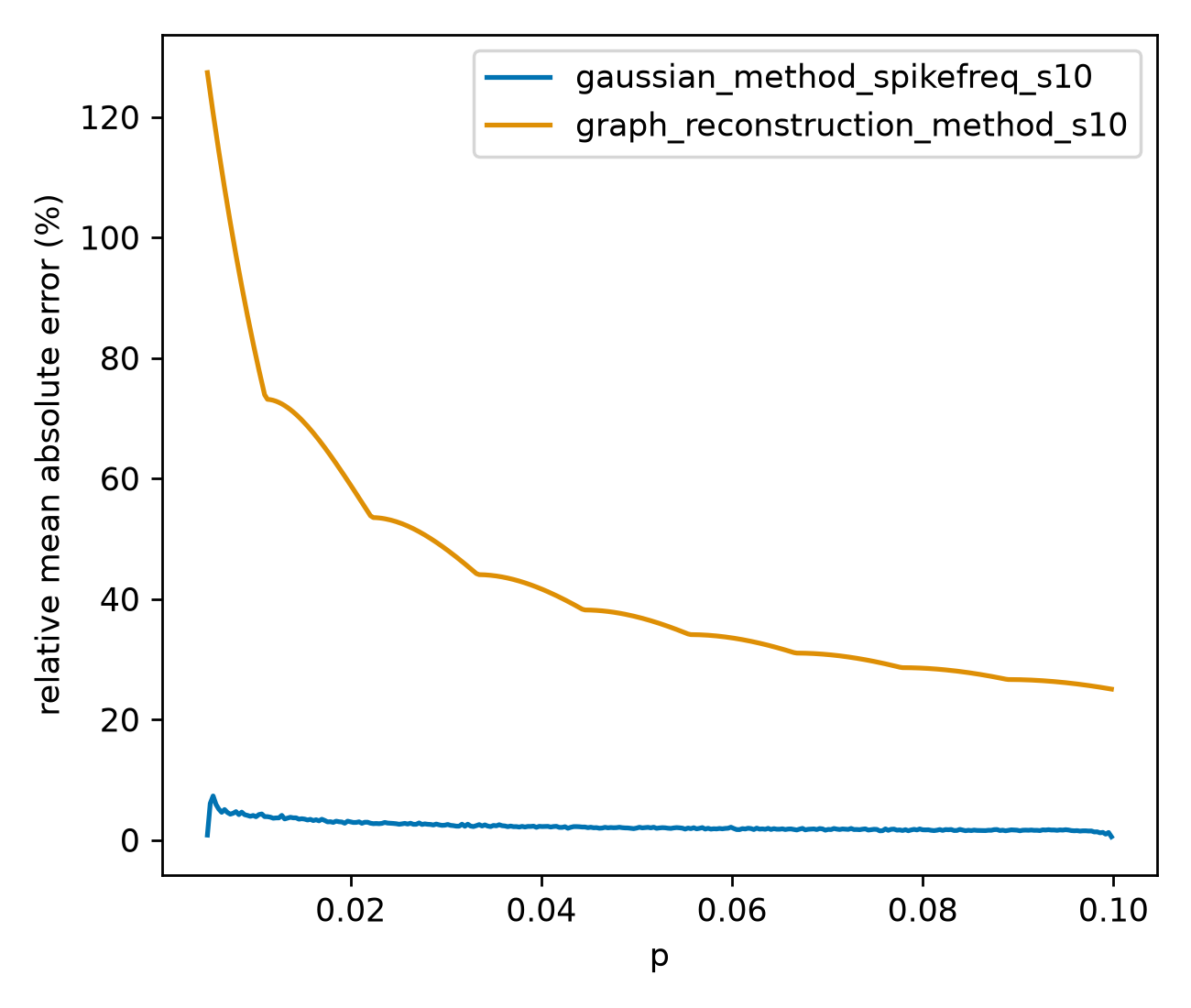}
  \caption{Mean absolute relative error}
  \label{fig:compare_methods_gr:mae}
\end{subfigure}
\begin{subfigure}{.49\textwidth}
  \centering
  \includegraphics[width=1\linewidth]{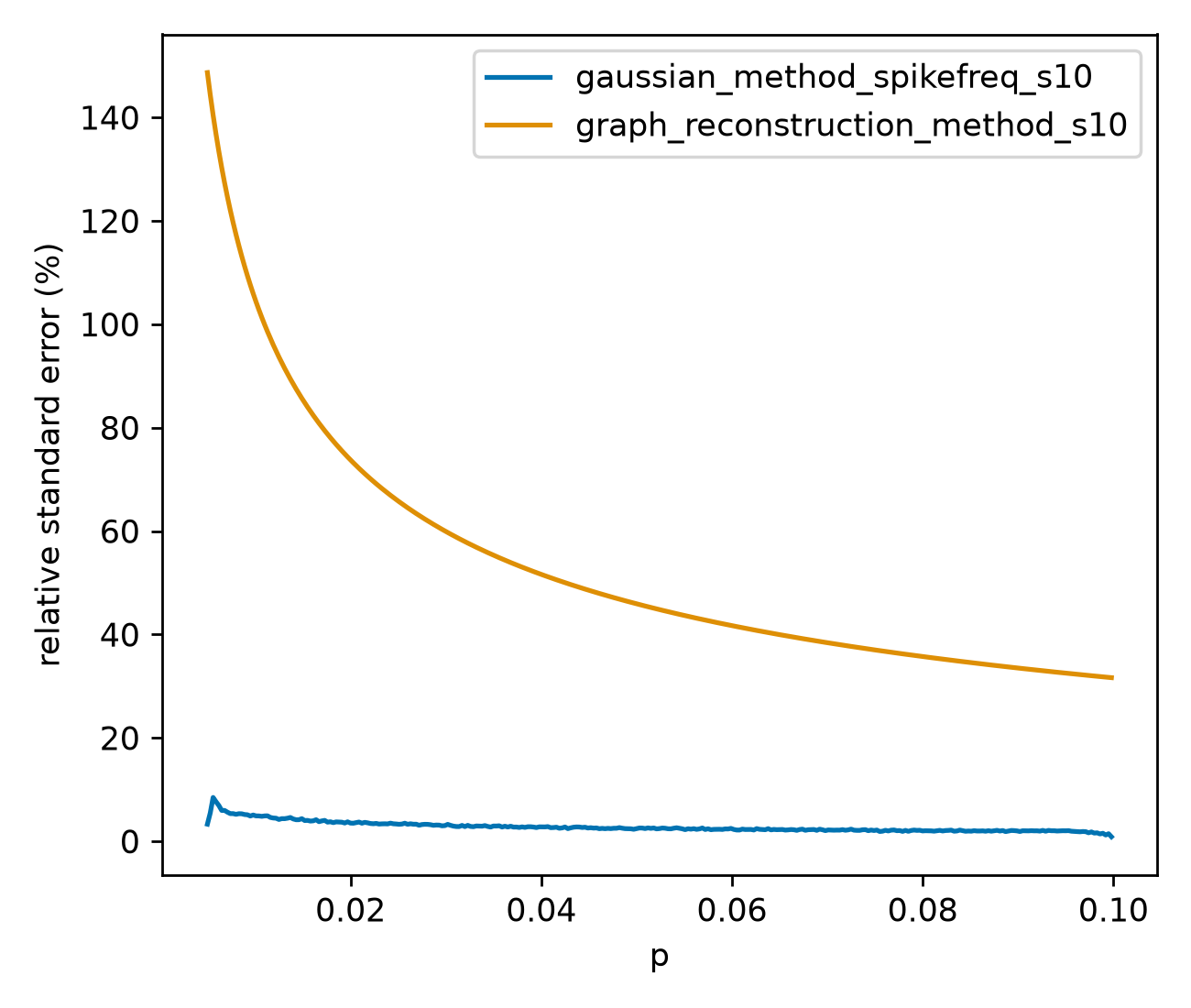}
  \caption{Relative standard error}
  \label{fig:compare_methods_gr:std}
\end{subfigure}
\caption{Relative mean absolute error (a) and standard error (b) of the estimation of $p$ for our method (with $\varphi$ being the numbe of spikes, and parametric Gaussian likelihood) and for the optimal graph reconstruction method.}
\label{fig:compare_methods_gr}
\end{figure}

This optimal mean absolute error and standard error are about 20 times larger than our best method, as shown in Fig.~\ref{fig:compare_methods_gr}.

\subsection{Removing reciprocal connections}
\begin{figure}[]
\centering
\begin{subfigure}{.49\textwidth}
  \centering
  \includegraphics[width=1\linewidth]{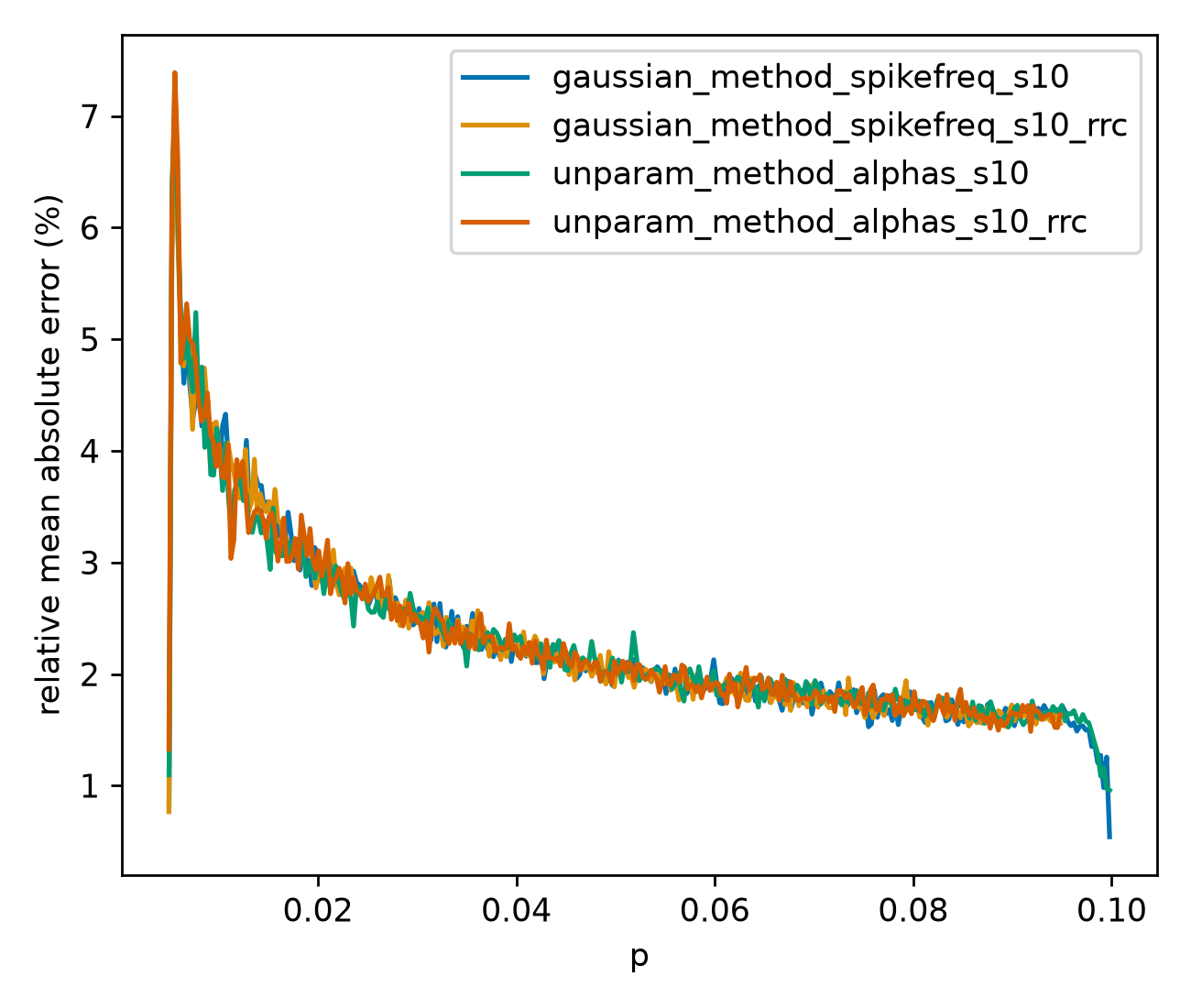}
  \caption{Mean absolute relative error}
  \label{fig:compare_methods_rrc:mae}
\end{subfigure}
\begin{subfigure}{.49\textwidth}
  \centering
  \includegraphics[width=1\linewidth]{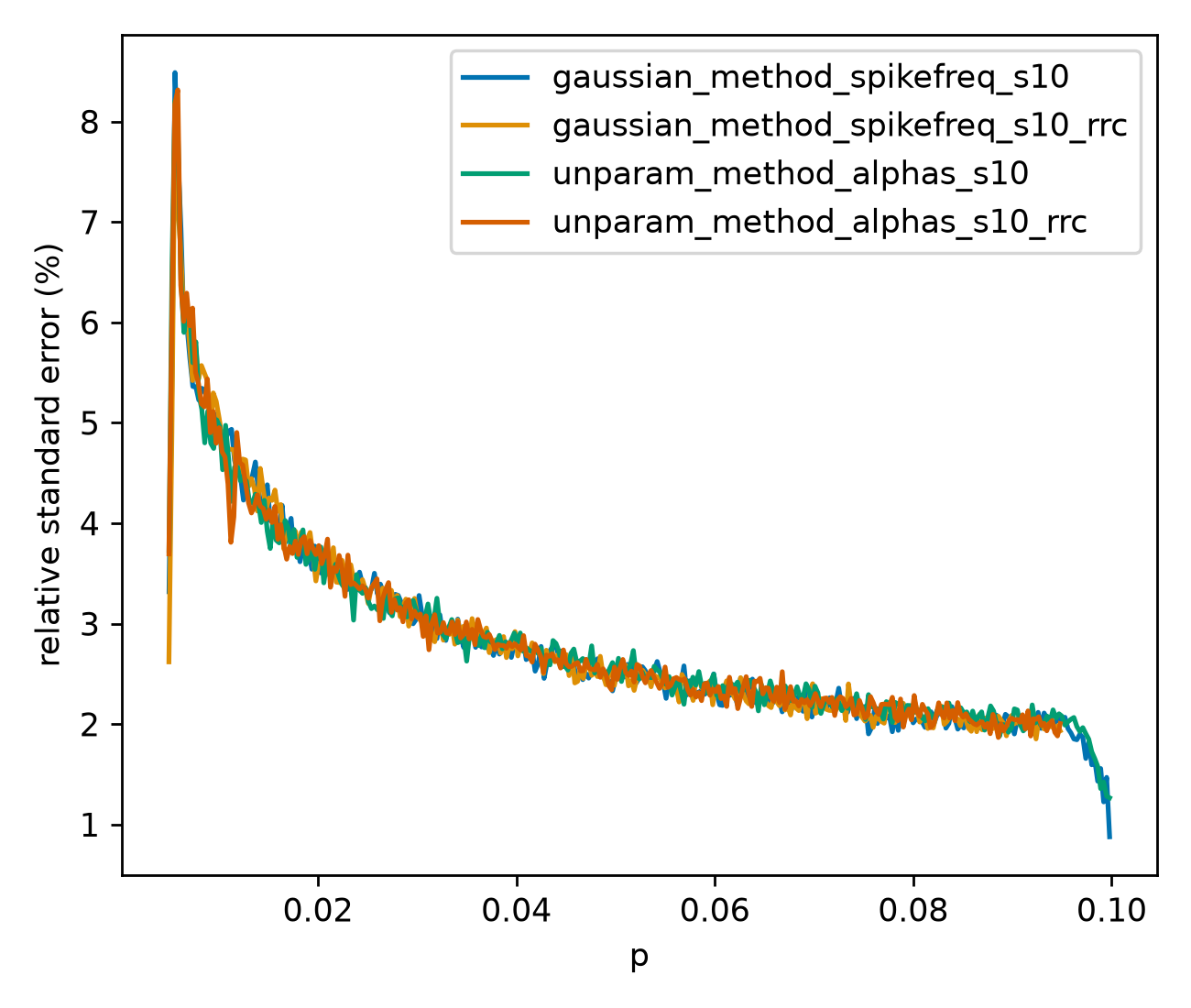}
  \caption{Relative standard error}
  \label{fig:compare_methods_rrc:std}
\end{subfigure}
\caption{Relative mean absolute error (a) and standard error (b) of the estimation of $p$ for variants of our method with (\texttt{rrc} suffixes in the legend) and without removed reciprocal connections.}
\label{fig:compare_methods_rrc}
\end{figure}

Given a neural network, a reciprocal connection is a couple of neuron $(i, j)$ where there is both a connection from $i$ to $j$ and a connection from $j$ to $i$. For such a couple, those reciprocal connections can be removed by randomly choosing one of the two connections and removing it. The connection proportion between the neurons is now $p' = p-p^2/2$. We ran our method with a neural network having reciprocal connections removed. Fig.~\ref{fig:compare_methods_rrc} shows that our method stills correctly estimates the relevant connection probability, $p'$, but does not exhibit signs that the fitted model, that does not exclude reciprocal connections, is not valid.

\subsection{Confidence intervals}
\label{sec:confidence-intervals}
\begin{figure}[]
\centering
\includegraphics[scale=0.7]{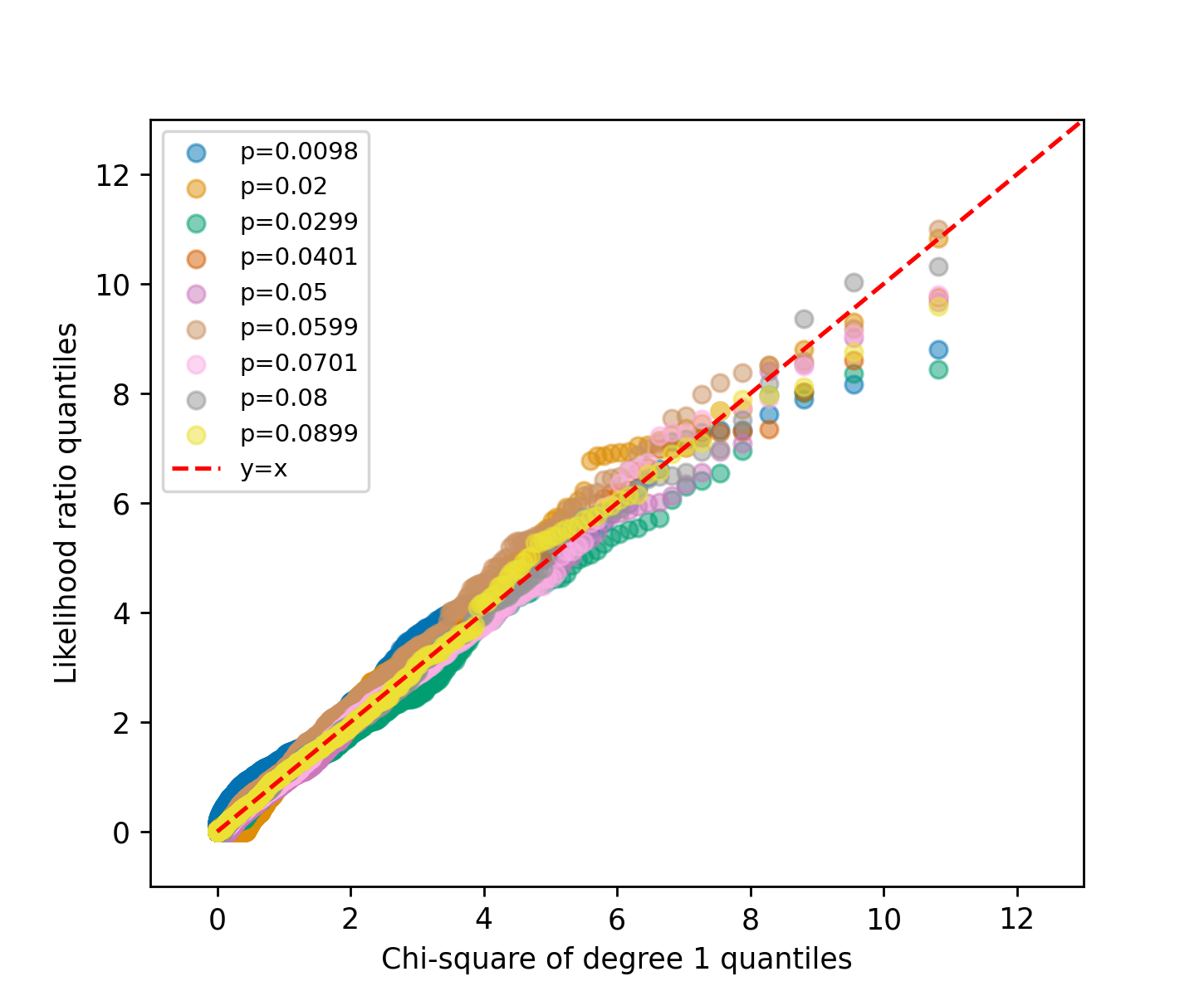}
\caption{Q-Q plot of estimated ``likelihood ratio'' (Eq.~\ref{eq:ratio-stat}) against Chi-squared distribution, for different values of $p$. The observed samples are frequency of spikes. The estimation method is Gaussian parametric.}
\label{fig:lr_qqplot}
\end{figure}

\begin{figure}[]
\centering
\includegraphics[scale=0.7]{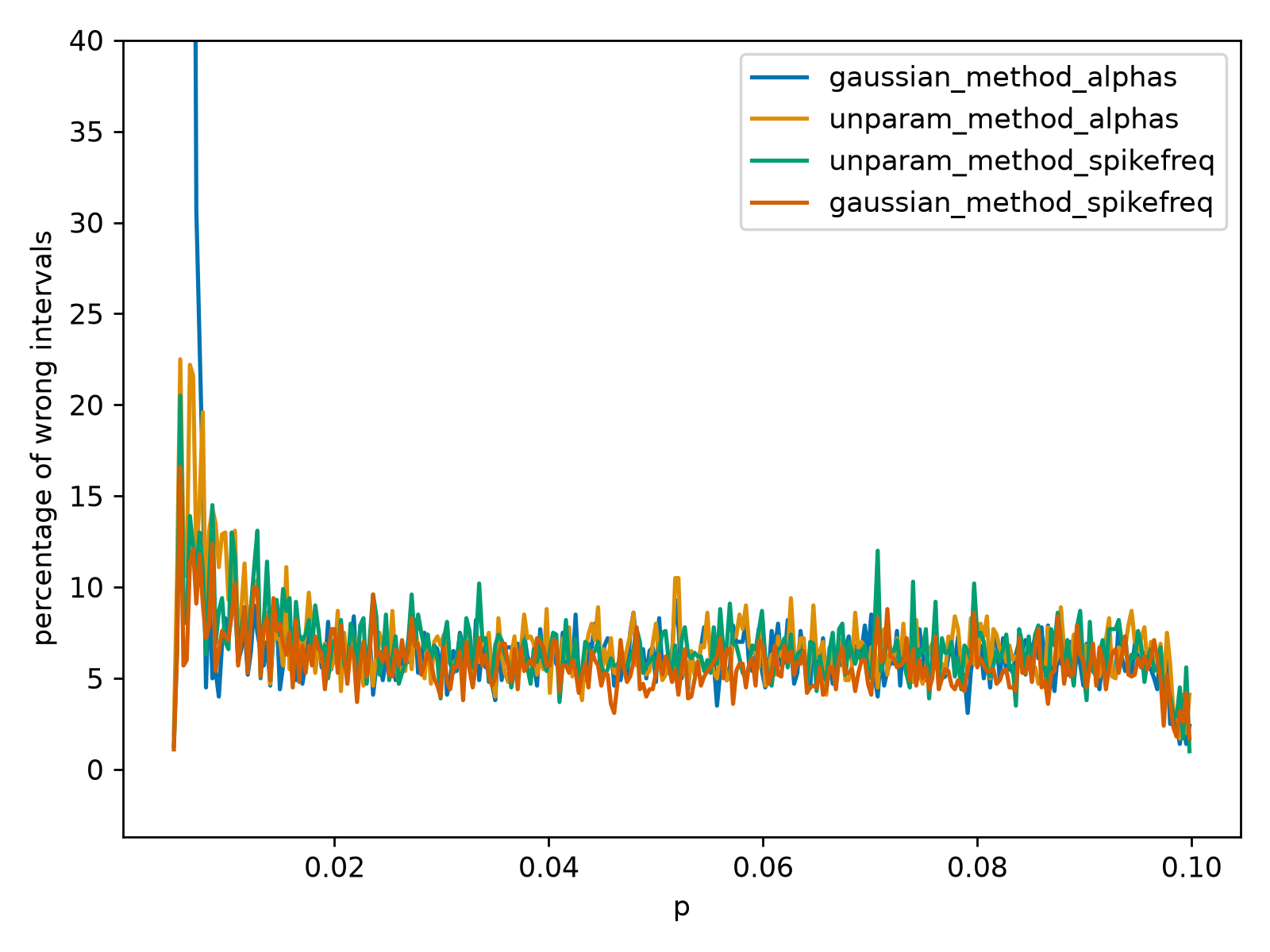}
\caption{Percentage of $95\%$ confidence intervals that do \textbf{not} contain the real value of $p$, for different estimation methods. One degenerate value of $90\%$ has been cropped for readability.}
\label{fig:compare_ci}
\end{figure}

Our method can also provide confidence interval for the value of the parameter $p$. We make the following assumption concerning the ``likelihood ratio'' statistic \cite[Sec. 8.2.1 and Theorem 10.3.3]{casella.berger:2002}:
\begin{equation}\label{eq:ratio-stat}
  D = -2\left(\log \widehat{\mathbb{P}}_{p_0}\{T(\mathbf{Y})=T(\mathbf{y})\} - \log \widehat{\mathbb{P}}_{\widehat{p}}\{T(\mathbf{Y})=T(\mathbf{y})\}\right) \approx \chi^2_{(1)}\, ,
\end{equation}
where $p_0$ is the actual $p$ value. Fig.~\ref{fig:lr_qqplot} shows the validity of this approximation for one of the variants of our method. With this assumption, an approximated $100\alpha\%$ confidence interval is the interval of value of $p$ such that
\[\log \widehat{\mathbb{P}}_{p}\{T(\mathbf{Y})=T(\mathbf{y})\} \geq \log \widehat{\mathbb{P}}_{\hat{p}}\{T(\mathbf{Y})=T(\mathbf{y})\} - \frac{x_{\alpha}}{2}\]
where $x_{\alpha}$ is the $\alpha$ quantile of the Chi-squared law, therefore is such that \[\mathbb{P}(D \leq x_{\alpha}) \approx \alpha.\]
Fig.~\ref{fig:compare_ci} shows that this method yields a good approximation of a $95\%$ confidence interval.

\section{Conclusions and perspectives}
\label{seq:conclusions}
The approach illustrated in this article was motivated by the need to face a \emph{fundamental limitation of extracellular recordings} of neural networks---only a tiny fraction of the neurons making an actual network are recorded---, and by the desire to address \emph{a question whose answer should be reproducible upon experiment repetitions}: what are the features of the random graph model from which the actual network observed is assumed to be a realization? When addressing questions about the graph structure, the ``fundamental limitation'' leads either to the reconstruction of the sub-graph formed by the recorded neurons---but that implies \emph{de facto} ``throwing away'' most of the recorded data \cite{Bartho_2004}---, or to the estimation of a \emph{functional graph} that is reproducible neither upon changes of the recording conditions (like the presentation of a stimulation), nor upon experiment repetitions. Our focus on the underlying random graph model parameters is an attempt to address the above fundamental limitation.

Our way of addressing the fundamental limitation issue is to use a numerical model of the \emph{whole} network to generate, by simulations, artificial data, before sampling the network activity in a way mimicking what is done in actual experiments. Considering the amount of knowledge we now have both on individual neuron morphology and on neuronal physiology/biophysics, we think  that simulating a credible numerical model of an entire network, at least when the latter is small enough, is easily doable. What constitutes a ``credible numerical model'' is clearly a hotly debated topic in the computational neuroscience community. Our choices of a simple stochastic model have been discussed elsewhere \cite[Chap. 1 and Appendix A]{galves.locherbach.pouzat:2024}; a different choice based on detailed biophysical models would be possible within our framework, the resulting computational cost would only be much higher. Since we are still at the proof of principle stage, we have opted for a toy model---meaning that this model should not be taken as a serious candidate for comparison with actual data---depending on a single parameter: the connection probability, $p$, of a directed Erdős-Rényi model (Sec.~\ref{sec:directed-ER-model}).

Having decided to sample the activities of few neurons from a whole network simulation, leaves open the question of how these simulated spike trains should be compared to the actual ones in order to adjust our model parameter $p$. Since comparing the spike sequences was clearly hopeless (Sec.~\ref{sec:an-intractable-likelihood}), we have followed the classical ABC/SBI path \cite{Sisson_2018} using a summary statistic computed from the spike sequences (Sec.~\ref{general_method:transformation} and \ref{general_method:particular_case}). To our surprise, the simplest such statistic, the number of spikes generated by each neuron, or their spiking rate   (Sec.~\ref{appendix:statistic}), turned out to give the best results (Sec.~\ref{sec:inference} and \ref{results:state_of_art}). In addition, we have shown that our method yields relevant confidence intervals (Sec.~\ref{sec:confidence-intervals}). If the role of the spiking rate distribution is confirmed in more realistic settings, this would constitute a very important result, since this quantity is easily measurable in experiments, and is almost systematically reported in neurophysiological articles.

This work is obviously only a first (tiny) step. We need to refine our numerical model to make its simulated activity comparable to actual data from our ``target'' network: the locust first olfactory relay \cite{Laurent_1996}. This implies including another neuronal type, inhibitory neurons as well as using neuro-anatomical data for proposing a more credible random graph model. On the inference side, that will require using more than a single scalar summary statistic; there is clearly some work to be done, but we are not navigating uncharted territories, since these many summary statistics are a usual feature of commonly implemented ABC/SBI methods \cite{Sisson_2018,Cranmer_2020}.

We also hope that this simulation study will encourage more theoretically oriented colleagues to obtain analytical results relative to the distribution of spiking frequencies in our setting:  a directed Erdős-Rényi model for the structure and a Galves-Löcherbach model for the dynamics. Such results would allow us to skip the ``heavy'' simulations leading to the set of estimated sampling distributions $\left\{\widehat{\mathbb{P}}_p\left\{\varphi(Y)=\varphi(y)\right\}\right\}_{p \in \mathcal{P}}$ (Sec.~\ref{model:sbi}).

\appendix

\section{Our heuristic approach to find statistics on spike trains}
\label{appendix:statistic}
In order to find a relevant statistic, that correctly discriminates the values of $p$, and leads to a simple probability density function, we propose to use the inter-spike interval (ISI) distribution of a neuron. Fig.~\ref{fig:isi_gamma_infer} shows an example of such a distribution. This is a compression of a neuron's spike train, and we try and show that it does not loose that much information.

\subsection{Dimension reduction}
\label{statistic:dim_red}
\begin{figure}[]
\centering
\includegraphics[scale=0.75]{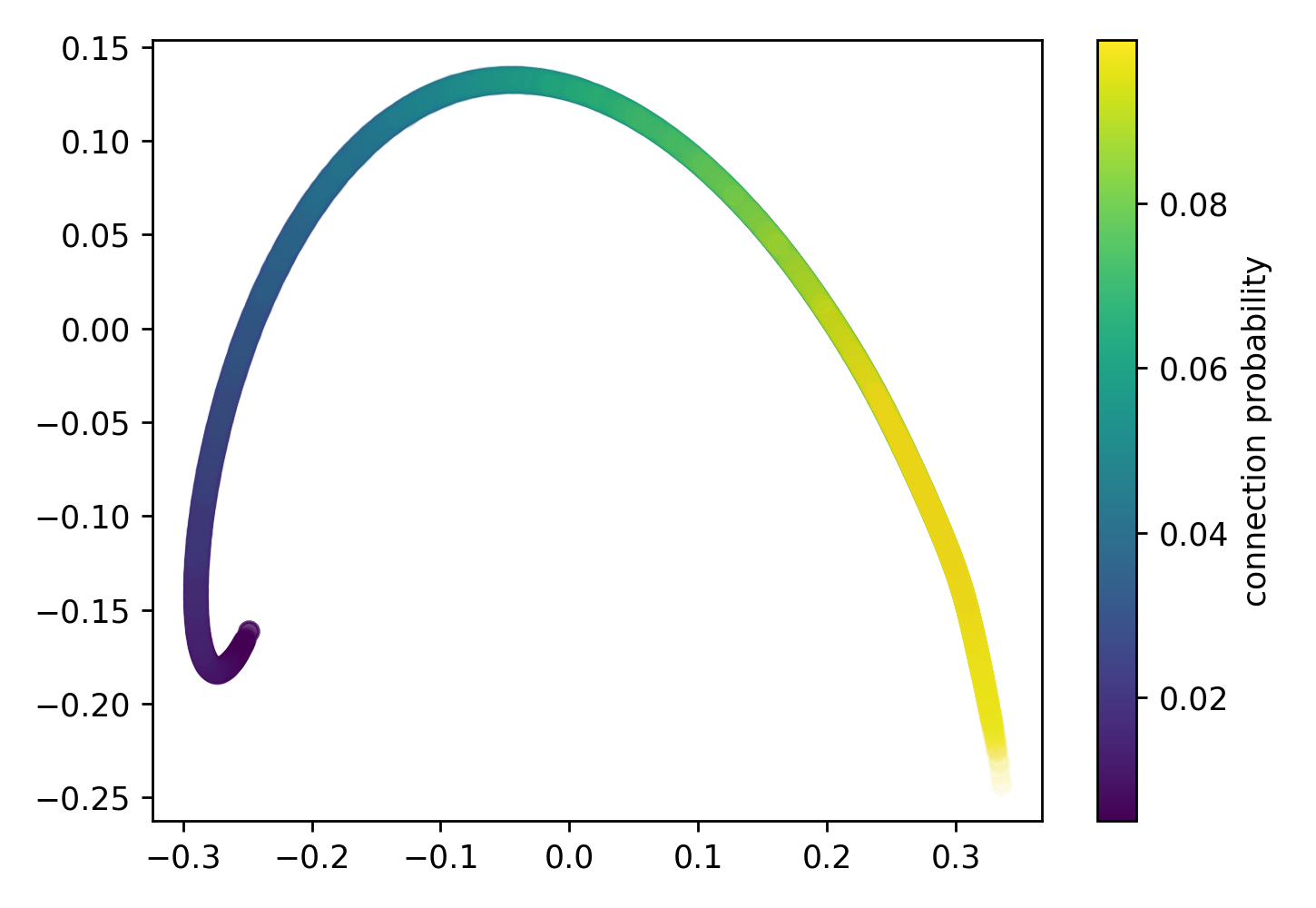}
\caption{Projection onto a 2 dimensional space of 96 000 ISI distributions coming from 96 different values of $p$ in $\mathcal{P}$.}
\label{fig:pca2d_proj}
\end{figure}

We first verify that there is no interspike interval larger than 100 time steps, as seen in Fig.~\ref{fig:isi_gamma_infer}. Then, an ISI distribution can be seen as a vector in $\mathbb{R}^{100}$ on which we perform dimension reduction. We simulate $317$ different graphs of size $n=1000$ neurons, one for each value of $p$ in  $\mathcal{P}' = \{0.005 + 0.0003\cdot i,\ 0 \leq i \leq 316\}$. This gives $317\ 000$ different ISI distributions. Performing PCA on this $317\ 000$ observations, and keeping the first two components, yields a curve. This curve is moreover quite well parameterized by the value of $p$ as can be seen in Fig.~\ref{fig:pca2d_proj}. This implies that there must be a way to project the ISI distributions onto a one dimensional space, such that it is possible to distinguish the associated value of $p$.

\subsection{ISI distribution summarized as its Gamma parameters}
\label{statistic:gamma_param}
\begin{figure}[]
\centering
9\includegraphics[scale=0.75]{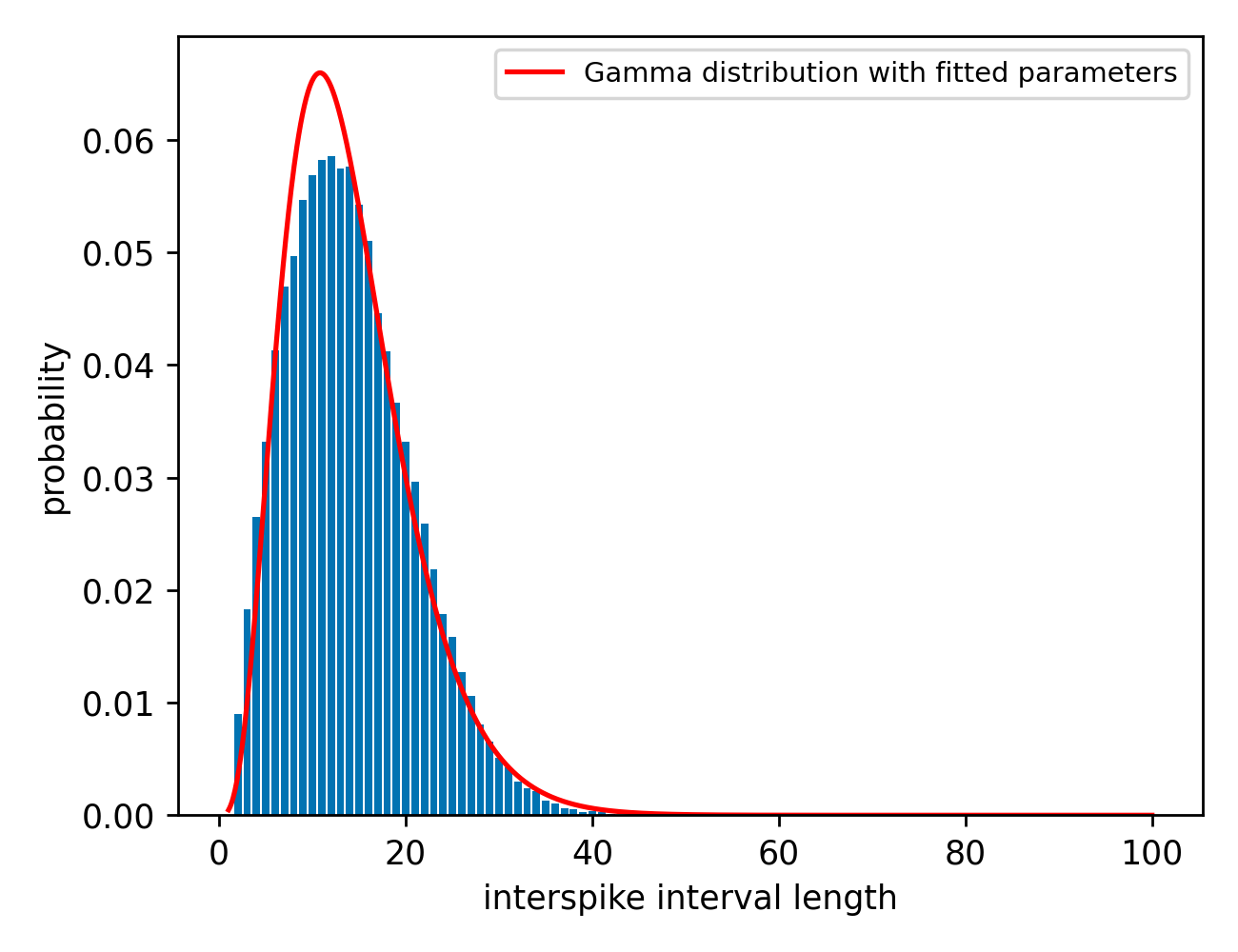}
\caption{Estimated density of the ISI distribution of a neuro in a graph with 1000 neurons, with a connection probability of $0.0122$ and a simulation time period of $10^6$ time steps. In red is the Gamma density with parameters inferred from the moment method.}
\label{fig:isi_gamma_infer}
\end{figure}

A way to summarize the ISI distributions with one or a few parameters, is to see it as a probability density function. Fig.~\ref{fig:isi_gamma_infer} shows that an ISI distribution ``looks like'' some parametric continuous probability density function. In particular, as it can be seen in Fig.~\ref{fig:isi_gamma_infer}, it is quite close to the Gamma distribution.

The easiest, and best, up to our experiments, distribution to use is the Gamma distribution. It is parameterized by a shape parameter, denoted as $\alpha$ and a rate parameter, denoted as $\beta$. Given a random variate $X$ distributed according to an ISI distribution, its parameters are given by : $$\alpha = \frac{\mathbb{E}[X]^2}{\mathrm{Var}[X]}, \quad \beta =  \frac{\mathbb{E}[X]}{\mathrm{Var}[X]}. $$ Thus, they can be estimated using the moment method, since our goal is to exhibit a statistic that is both simple and interpretable.

\begin{figure}[]
\centering
\includegraphics[scale=0.75]{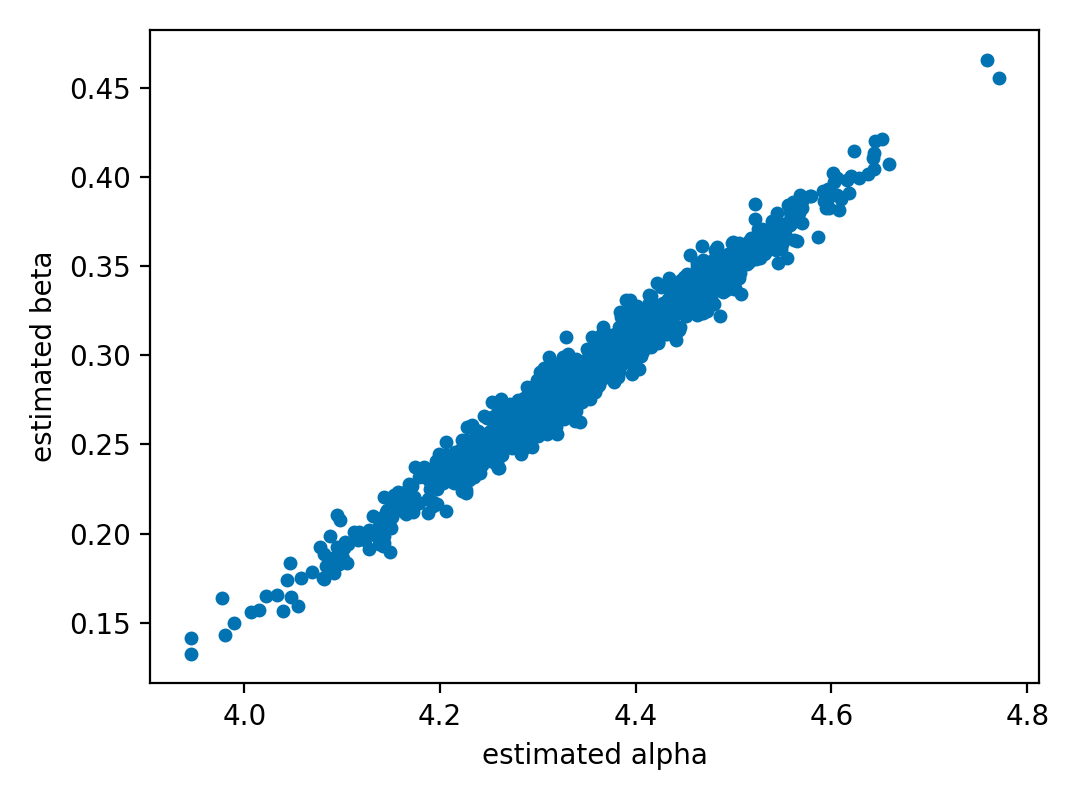}
\caption{Estimated $\alpha$ parameter versus estimated $\beta$ parameter of the Gamma distribution fitted to each ISI distribution, corresponding to each spike trains from a graph generated with connection probability $p = 0.0122$ and $n = 1000$ neurons.}
\label{fig:gamma_param}
\end{figure}

For a given value of $p$,  we plot the estimated $\beta$ against the estimated $\alpha$ for each Gamma distribution fitted to each ISI distribution corresponding to each spike train from one given  neural network and its time evolution. The points gather closely to a straight line, as shown in Fig.~\ref{fig:gamma_param}. This shows a clear correlation between the two parameters, implying that using only of the two parameters is enough to describe the ISI distribution.

\begin{figure}[]
\centering
\includegraphics[scale=0.75]{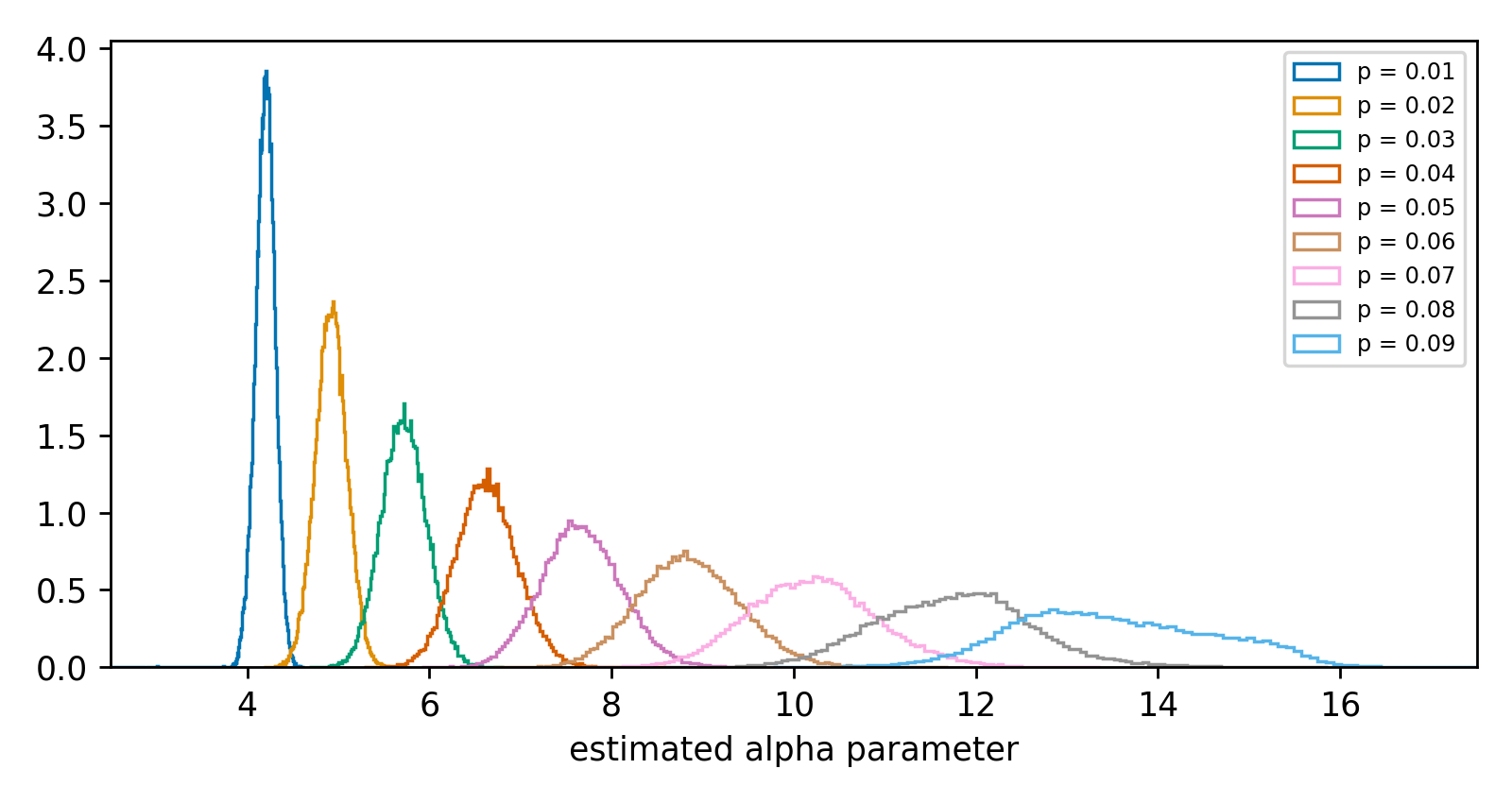}
\caption{Distributions of the estimated $\alpha$ parameters for different values of $p$. For each $p$, 50 graphs of 1000 neurons were generated and simulated, from which the $1000$ spike trains were extracted, and a total of $50000$ ISI distributions were parametrically estimated, yielding $50000$ estimated $\alpha$ parameters for each $p$.}
\label{fig:hist_alpha}
\end{figure}

In the end, our empirical result shows that the estimated $\alpha$ parameter is indeed a good statistic to distinguish between the values of $p$ as we can see in the Fig.~\ref{fig:hist_alpha}.

\subsection{ISI distribution summed up as its mean}
\label{statistic:spikefreq}
\begin{figure}[]
    \centering
  \includegraphics[scale=0.75]{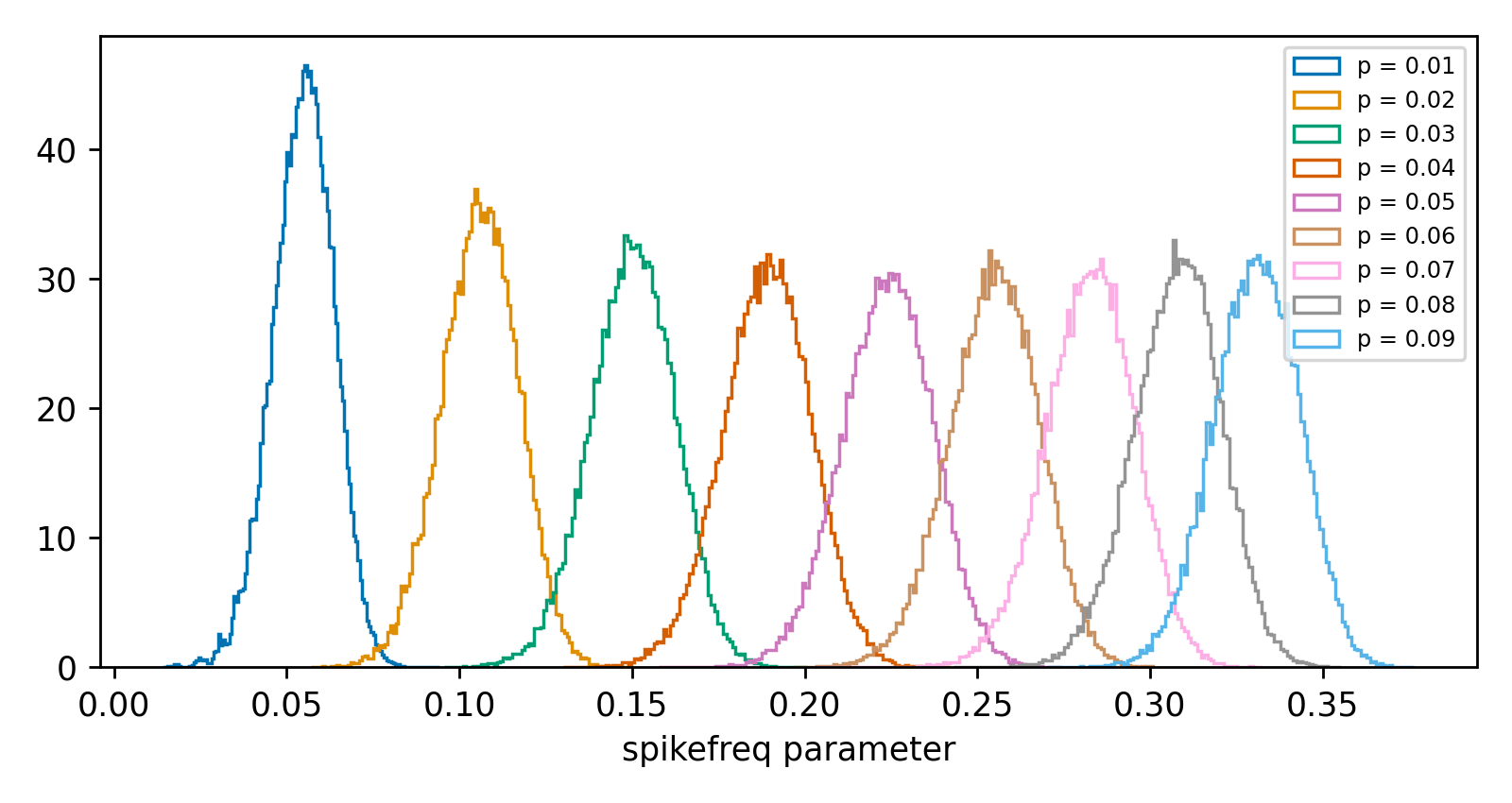}
  \caption{Distributions of the spikes frequencies for different values of $p$. For each $p$, 50 graphs of 1000 neurons were generated and simulated, from which the $1000$ spike trains were extracted, yielding a total of $50000$ spike frequencies for each $p$.}
  \label{fig:hist_spikefreq}
\end{figure}

Fig.~\ref{fig:gamma_param} suggests that, for a given $p$, the ratio $\widehat{ \alpha} / \widehat{\beta}$, which is exactly the estimated mean of the ISI distribution, remains approximately constant. This suggests that we could use the estimated mean of the ISI distribution as the statistic on the spike trains. This statistic also discriminates correctly between the values of $p$, as shown in Fig.~\ref{fig:hist_spikefreq}.
\begin{remark}
The average interspike interval length is, in fact, the number of time steps of the simulation divided by the frequency of spikes of the neuron at hand during the whole time period. In practice, this is more convenient to use the inverse of this statistic, \textit{i.e.} the frequency of spikes.
\end{remark}

\section{Maximizing \texorpdfstring{$\widehat{\mathbb{P}}_p\left\{T(\mathbf{Y})=T(\mathbf{y})\right\}$}{the estimated likelihood} with respect to \texorpdfstring{$p$}{p}}
\label{appendix:maximum_estimation}
\begin{figure}[]
\centering
\begin{subfigure}{.49\textwidth}
  \centering
  \includegraphics[width=1\linewidth]{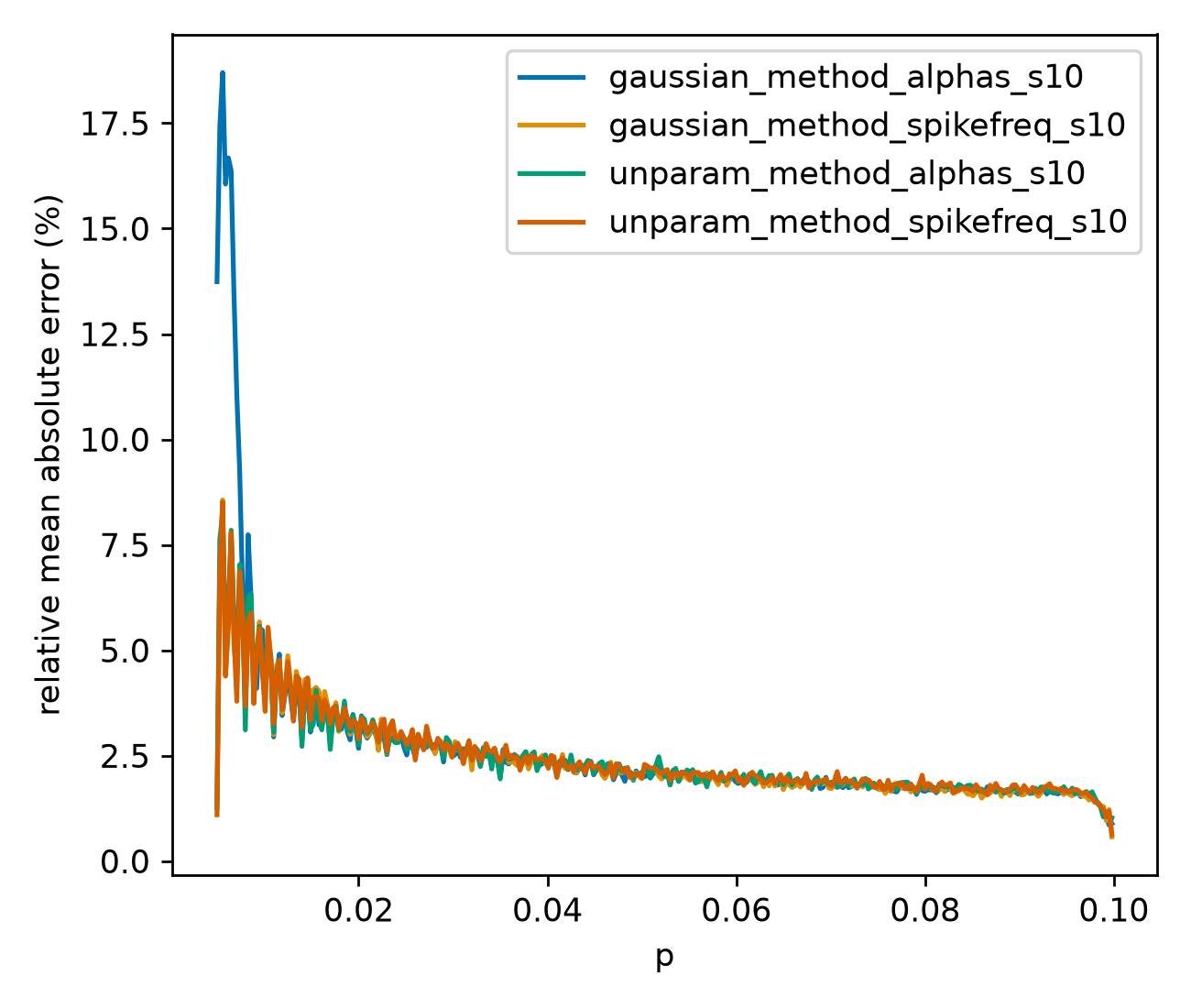}
  \caption{Discrete maximum estimation}
  \label{fig:compare_continuous:discrete}
\end{subfigure}
\begin{subfigure}{.49\textwidth}
  \centering
    \includegraphics[width=1\linewidth]{compare_methods_continuous_estimation_mean_absolute_error.png}
  \caption{Continuous maximum estimation}
  \label{fig:compare_continuous:continuous}
\end{subfigure}
\caption{Relative mean absolute error of our method, with discrete (a) or continuous (b) maximum estimation}
\label{fig:compare_continuous}
\end{figure}

Our method requires to find the maximum of $\widehat{\mathbb{P}}_p\left\{T(\mathbf{Y})=T(\mathbf{y})\right\}$ with respect to $p$, with $\mathbf{y}$ fixed.
This maximization is first performed for $p \in \mathcal{P}$, a discrete grid, as described in Sec.~\ref{model:sbi}. This method clearly induces a bias as soon as the actual $p$ is not in the set $\mathcal{P}$. This can be seen as oscillations on Fig.~\ref{fig:compare_continuous:discrete}.

To overcome this issue, we implemented a second step yielding a continuous estimate of the maximum of the function using a quadratic interpolation between the grid values.
Let $\mathcal{P} = \{p_1, \dots, p_N\}$, and let $i_{\text{max}} \in \{1, \dots, N\}$ be such that $\widehat{\mathbb{P}}_p\left\{T(\mathbf{Y})=T(\mathbf{y})\right\}$ is maximized at $p=p_{i_{\text{max}}}$. 
We can then easily compute the quadratic polynomial that interpolate the 3 points $\left(p_{i_{\text{max}} -1}, \widehat{\mathbb{P}}_{p_{i_{\text{max}} -1}}\left\{T(\mathbf{Y})=T(\mathbf{y})\right\}\right)$, $\left(p_{i_{\text{max}}}, \widehat{\mathbb{P}}_{p_{i_{\text{max}}}}\left\{T(\mathbf{Y})=T(\mathbf{y})\right\}\right)$ and $\left(p_{i_{\text{max}}+1}, \widehat{\mathbb{P}}_{p_{i_{\text{max}} +1}}\left\{T(\mathbf{Y})=T(\mathbf{y})\right\}\right)$.
This polynomial is necessarily concave, and the value maximizing this quadratic polynomial is our estimator $\widehat{p}$.
Lemma \ref{lemma:interpolate} gives an efficient algorithm to compute this maximizing value.

\begin{lemma}
    \label{lemma:interpolate}
    Let $a_1 = (x_1, y_1), a_2 = (x_2, y_2), a_3 = (x_3, y_3)$ be three points in $\mathbb{R}^2$, such that $x_1 < x_2 < x_3$, and $y_2 \geq y_1, y_2 \geq y_3$. 
    Then the value maximizing the quadratic polynomial interpolating $a_1, a_2$ and $a_3$ is 
    $$x_{\max} =  \frac{\Delta_1(x_2+x_3)  - \Delta_2(x_1+x_3) + \Delta_3(x_1+x_2)}{2(\Delta_1 - \Delta_2 + \Delta_3)}$$
    where $$\Delta_1 = y_1(x_3-x_2), \quad \Delta_2 = y_2(x_3-x_1), \quad \Delta_3 = y_3(x_2-x_1).$$
\end{lemma}

\begin{remark}
    Note that this step could boost the noises caused by the many approximations that were made. However, as seen in Fig.~\ref{fig:compare_continuous}, this does not impact our results. We believe that there is only very small noises, and the final composite likelihood is always ''reasonable''. However, this is important to note that it could not be the case anymore for more complex model.
\end{remark}

\section{Proofs}
\label{sec:proofs}
\subsection{Proof of Lemma 4.1}
If the neural network is generated as described in Sec. 3.3, $\widehat{N_c}$ follows a binomial distribution $B(s(s-1), p)$. The mean absolute error is then
\begin{align*}
  \mathbb{E}\big[|\widehat{p} - p|\big] &= \mathbb{E}\Bigg[\Big|\frac{\widehat{N_c}}{s(s-1)} - p\Big|\Bigg] \\
  &= \mathbb{E}\Bigg[\Big(\frac{\widehat{N_c}}{s(s-1)}-p\Big)\mathbf{1}_{\{\widehat{N_c}\geq s(s-1)p\}}\Bigg] + \mathbb{E}\Bigg[\Big(p-\frac{\widehat{N_c}}{s(s-1)}\Big)\mathbf{1}_{\{\widehat{N_c}<s(s-1)p\}}\Bigg] \\ 
                                        &= \mathbb{E}\Bigg[\Big(\frac{\widehat{N_c}}{s(s-1)}-p\Big)\Bigg] - \mathbb{E}\Bigg[\Big(\frac{\widehat{N_c}}{s(s-1)}-p\Big)\mathbf{1}_{\{\widehat{N_c}<s(s-1)p\}}\Bigg] + \mathbb{E}\Bigg[\Big(p-\frac{\widehat{N_c}}{s(s-1)}\Big)\mathbf{1}_{\{\widehat{N_c}<s(s-1)p\}}\Bigg] \\
                                        &= 2 \mathbb{E}\Bigg[\Big(p-\frac{\widehat{N_c}}{s(s-1)}\Big)\mathbf{1}_{\{\widehat{N_c}<s(s-1)p\}}\Bigg] \\
  &= 2\sum_{k=0}^{\lceil s(s-1)p\rceil-1} \binom{s(s-1)}{k} p^k(1-p)^{s(s-1) - k}\Big(p-\frac{k}{s(s-1)}\Big)
\end{align*}

The standard error is
\[\sigma\Big(\frac{N_c}{s(s-1)}\Big) = \frac{1}{s(s-1)}\sqrt{s(s-1)p(1-p)} = \sqrt{\frac{p(1-p)}{s(s-1)}}\,.\]

\subsection{Proof of Lemma B.2}
    Let $P(X) = aX^2 + bX + c$ be the polynomial interpolating $a_1, a_2$ and $a_3$, then $x_{\max} = -b/(2a)$. \\
    Let $A = (x_3-x_1)(x_3-x_2)(x_2-x_1)$.
    Using Lagrange basis, we get that 
    $$P(X) = y_1\frac{(X-x_2)(X-x_3)}{(x_1-x_2)(x_1-x_3)} + y_2\frac{(X-x_1)(X-x_3)}{(x_2-x_1)(x_2-x_3)} + y_3\frac{(X-x_1)(X-x_2)}{(x_3-x_1)(x_3-x_2)}$$ 
    $$ = \frac{1}{A}\Big(y_1(x_3-x_2)\big(X^2 - (x_2+x_3)X + c_1\big) - y_2(x_3-x_1)\big(X^2 - (x_3+x_1)X + c_2\big) + y_3(x_2-x_1)\big(X^2 - (x_2+x_1)X + c_3\big)\Big)$$
    $$ = X^2\cdot\Big(\frac{y_1(x_3-x_2) - y_2(x_3-x_1) + y_3(x_2-x_1)}{A}\Big) + X\cdot\Big(\frac{y_2(x_3^2-x_1^2) - y_1(x_3^2-x_2^2) - y_3(x_2^2-x_1^2)}{A}\Big) + c_4.$$
    Then $$x_{\max} = \frac{ y_1(x_3^2-x_2^2) - y_2(x_3^2-x_1^2) + y_3(x_2^2-x_1^2)}{2(y_1(x_3-x_2) - y_2(x_3-x_1) + y_3(x_2-x_1))} = \frac{\Delta_1(x_2+x_3) - \Delta_2(x_1+x_3) + \Delta_3(x_1+x_2)}{2(\Delta_1 - \Delta_2 + \Delta_3)}.$$

\section{Discussion on the choice of \texorpdfstring{$K$}{K}}
In this section, we discuss the role of the number $K$ (see Sec.~\ref{model:sbi}) and justify the choice we made for its value.

The role of $K$ is to eliminate the variability, due to the random nature of the network on the distribution of the summary statistic under the parameter $p$. Our hope is that by repeating independent realization of the simulation, and by collecting $1000$ values of the summary statistic for each simulation, we are very close to estimate the true distribution $\mathbb{P}_p\left\{\varphi(Y) = \varphi(y)\right\}$. 

\begin{figure}[]
\centering
\begin{subfigure}{.49\textwidth}
  \centering
  \includegraphics[width=1\linewidth]{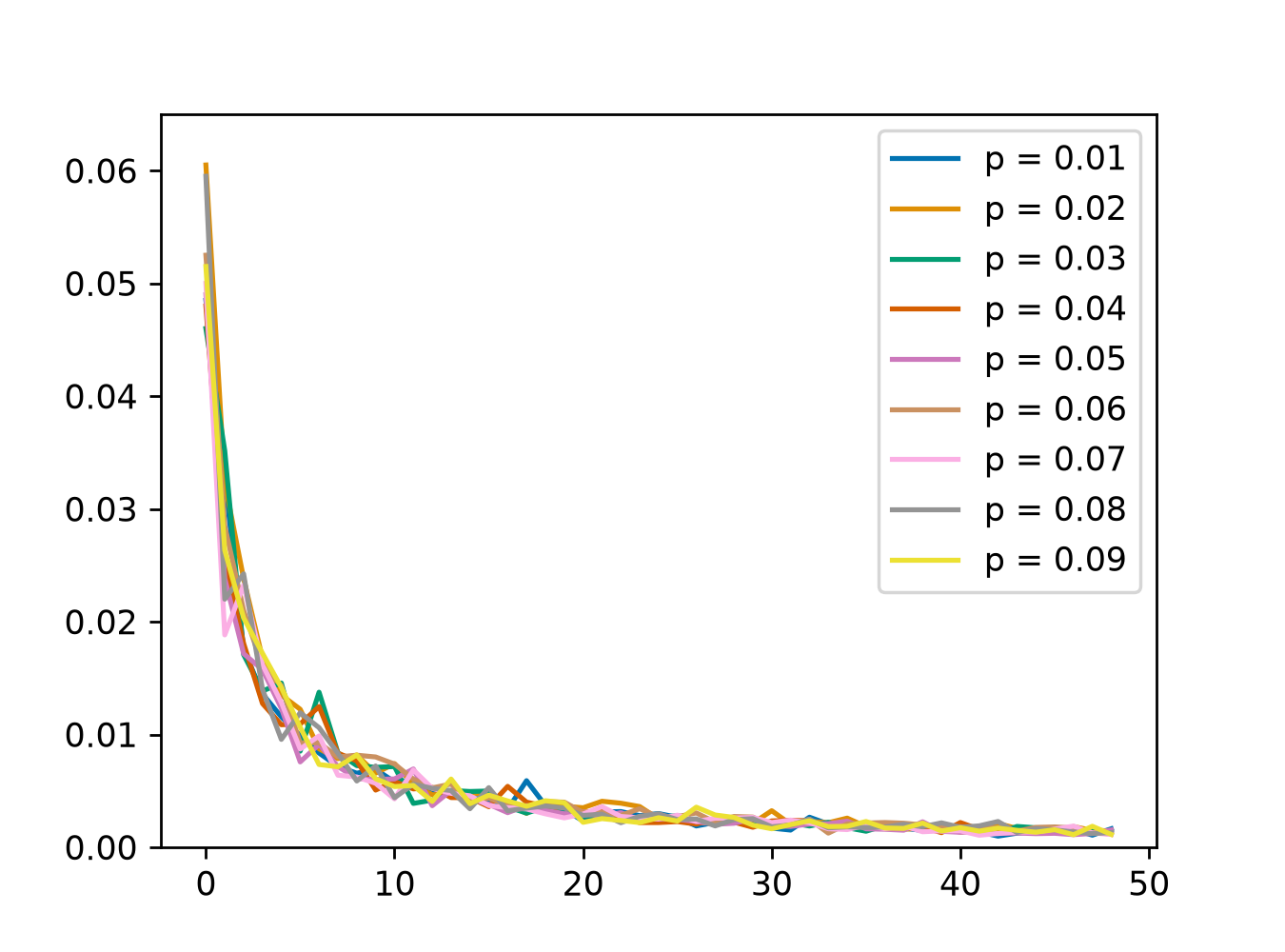}
  \caption{$\varphi$ is the frequency of spikes}
  \label{fig:convergence_K:spikefreq}
\end{subfigure}
\begin{subfigure}{.49\textwidth}
  \centering
    \includegraphics[width=1\linewidth]{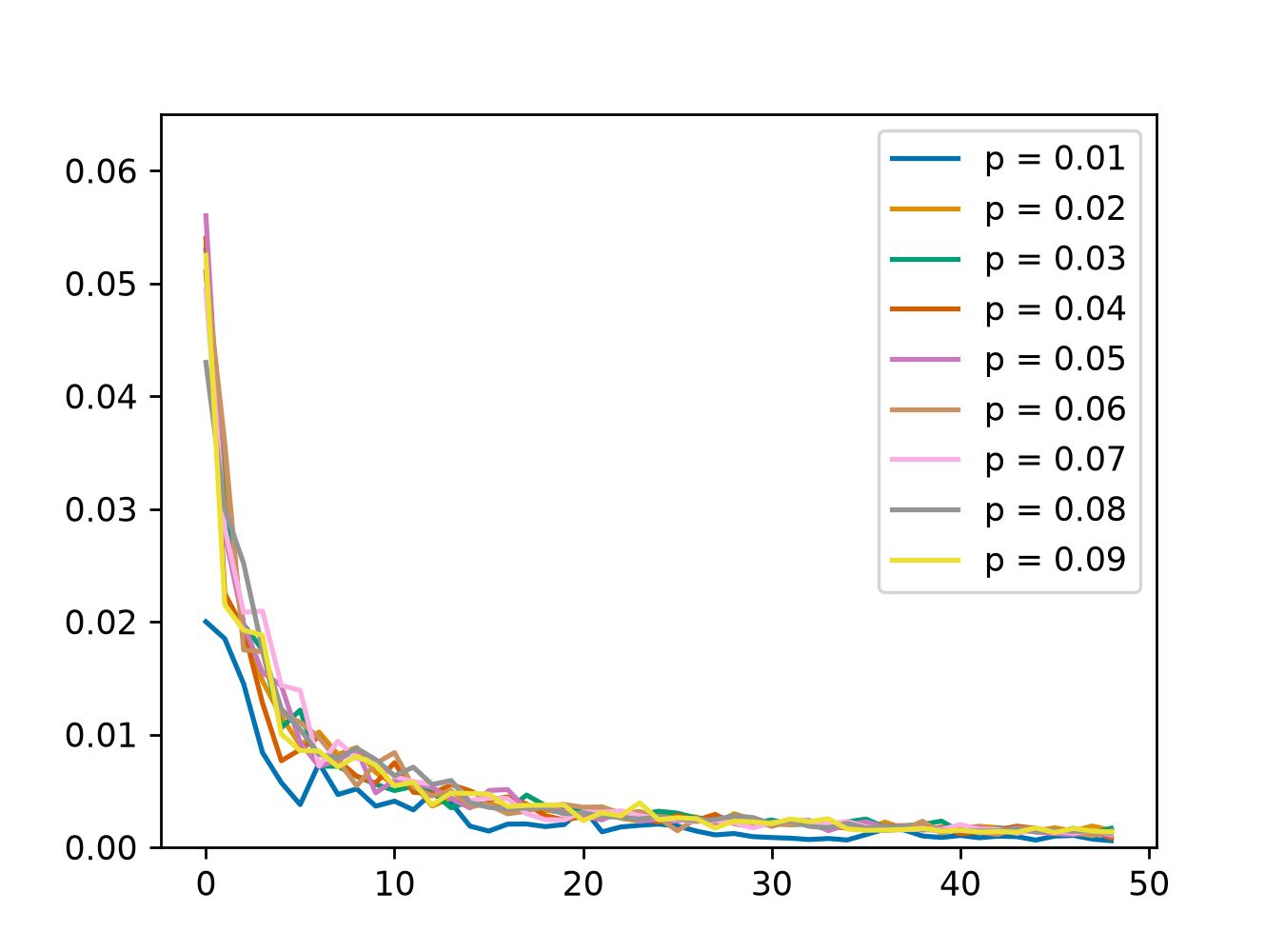}
  \caption{$\varphi$ is the estimated alpha parameter}
  \label{fig:convergence_K:alphas}
\end{subfigure}
\caption{Evolution as a function of $K$ of the total variation distance between the estimation of the distribution $\mathbb{P}_p\left\{\varphi(Y) = \varphi(y)\right\}$ with $K$ simulation, and $(K+1)$ simulation.}
\label{fig:convergence_K}
\end{figure}

To justify our choice of $K=50$, we can check that at this value, the estimation $\hat{\mathbb{P}}_p\left\{\varphi(Y) = \varphi(y)\right\}$ has already converged to the real distribution. To check this, we progressively increase the value of $K$, and compute the total variation distance between two consecutives estimation. 
In other terms we compute and plot this function $$K \mapsto d_{\text{TV}}\left(\hat{\mathbb{P}}^{(K)}_p\left\{\varphi(Y) = \varphi(y)\right\}, \hat{\mathbb{P}}^{(K+1)}_p\left\{\varphi(Y) = \varphi(y)\right\}\right).$$
When this function enough to zero (typically $<10^{-2}$), this means that increasing $K$ only lead to a negligeable change in the estimation, thus this value of $K$ is big enough. Fig.~\ref{fig:convergence_K} shows that for different values of $p$, and for both choices of $\varphi$, the value $K=50$ is more than large enough for this setup.

Note that this value $K$ is a parameter of this inference framework, that should be again tune in different setups, in particular for more complex model or descriptive statistics.

\section{Some numerical details}
\label{sec:some-numerical-details}
Most of the mythical man-month effort involved in the production of this article does not appear obviously from the previous pages. A practical implementation of SBI methods requires efficient codes for simulations, as well as, for summary statistics computations. A preliminary version of these codes was written for the statistical software \texttt{R}, but it became quickly clear that faster---meaning compiled---codes were required. We therefore developed our codes from scratch in \texttt{Fortran}. We won't discuss the pros and cons of \texttt{Fortran} versus \texttt{C} or \texttt{C++}; we just mention that we can program in all these languages and that our choice is not due to an incapability of using alternatives. From a wider perspective, what really matters is the use of a compiled code; our concern for reproducibility ``in time'' also favors the use of such \emph{standardized} languages (the three mentioned languages are standardized).

The \texttt{Fortran} standard has been requiring a pseudo-random number generator (PRNG) since \texttt{Fortran 90}, but (wisely) leaves open the algorithm used by the latter. In order to get reproducible results regardless of the compiler used, we implemented the \texttt{xoshiro256++} generator of \cite{Blackman_2021}. Our implementation was tested against the authors' reference \texttt{C} implementation\footnote{\url{https://prng.di.unimi.it/xoshiro256plusplus.c}.}. All the ``heavy'' computation was done using these \texttt{Fortran} programs, while light analysis and figures generation was done with \texttt{Python}. All our codes and scripts carrying out the simulations, their analysis and the generation of the figures of this article are available from our dedicated \texttt{GitLab} repository \texttt{network-codes-in-fortran}\footnote{\url{https://gitlab.math.unistra.fr/christophe.pouzat/network-codes-in-fortran}.}.

\section*{Acknowledgments}
This article is dedicated to the memory of Antonio Galves, a brilliant mathematician whose curiosity could not be sated, and a very dear friend of the author (CP) who had the privilege to know him well.

Pierre Charitat was supported by a fellowship from Strasbourg University Interdisciplinary Thematic Institute: Research in Mathematics, Interactions and Applications (IRMIA++).

Ségolen Geffray and Christophe Pouzat were supported by an ANR grant: SIMBADNESTICOST ANR-22-CE45-0027. 

The authors would like to thank te two referees for their insightful comments.

\pagebreak
\bibliographystyle{plain}
\bibliography{../bibliography}

\end{document}